# Coronavirus SARS-CoV-2: Analysis of subgenomic mRNA transcription, 3CLpro and PL2pro protease cleavage sites and protein synthesis

Corresponding autor: Miguel Ramos-Pascual

Abstract

Coronaviruses have recently caused world-wide severe outbreaks: SARS (Severe Acute Respiratory Syndrome) in 2002 and MERS (Middle-East Respiratory Syndrome) in 2012. At the end of 2019, a new coronavirus outbreak appeared in Wuhan (China) seafood market as the first focus of infection, becoming a pandemics in 2020, spreading mainly into Europe and Asia [Zu et al 2020]. Although the virus family is well-known and symptoms are similar to other coronaviruses (fever, pneumonia, small pleural effusions), this specific virus type presents considerable differences, as higher transmission and mortality rates, being a challenge for diagnostic methods, treatments and vaccines.

Coronavirus.pro (SARS-CoV-2) App is a module of Virus.pro, a C++ application which simulates Coronavirus replication cycle. This software identifies virus types in short times and provides FASTA files of virus proteins, a list of RNA sequences (regulatory, packaging, transcription and translation) and secondary structures (stem-loops, helices, palindromes, mirrors), once the virus genome has been sequenced. The code is supported by other bio-informatics tools, such as Vienna RNA package, Varna software and ClustalW2.

Coronavirus.pro has identified 2019-nCoV virus as a beta-coronavirus more close related to SARS type than to MERS. However, it presents significant differences, such as the spike glycoprotein precursor, characteristic of this virus type, and the increased number of transcription regulating sequences (TRS), producing more subgenomic mRNAS and synthesizing more fusion proteins than SARS/MERS. This could be related with those severe health effects (toxicity) on host patients than other coronaviruses.

The software has identified a list of structural, non-structural and accessory proteins in 2019-nCoV virus genome similar to SARS and MERS. It has found also several ORF encoding some accessory proteins with unknown TRS (i.e. AP3b, AP9b, AP11, AP12 and AP14a/b). Furthermore, there is a subgenomic mRNA, the shortest with 374bp, which translates no proteins, specific only of SARS type virus. Finally, there are some accessory proteins AP2 in SARS/MERS and AP2a/b in 2019-nCoV, encoded before ORF1.2 and ORF1.4 respectively, which have not been previously reported.

2019-nCoV protein sequences have been compared with those from SARS and MERS. As 3CLpro (nsp5) and RdRp (nsp12) have >90% similarities with SARS, some antiviral drugs effective with SARS coronavirus, such as protease inhibitors or RNA-dependent RNA polymerase inhibitors could be also effective to this virus type. Nevertheless, further comparisons would be required, including other types of estimators.

These results are useful as a first step with other bio-informatics and pharmacological tools in order to develop diagnostic methods (real time RT-PCR or ELISA tests), new vaccines or antiviral drugs, which avoid virus replication in any stage: fusion inhibitors, RdRp inhibitors and PL2pro/3CLpro protease inhibitors.

Keywords: SARS-CoV, MERS-CoV, 2019-nCoV, Coronavirus, virus proteins, protease cleavage sites

## 1. Introduction

Coronaviruses (CoVs) are specific viruses that cause diseases in mammals and birds, including humans, with symptoms such as enteritis in bats, mice and pigs and upper respiratory malfunctions and potentially lethal respiratory infections in humans [Fehr and Perlman 2015] A large variety of coronaviruseses have been previously studied and analyzed. These viruses are responsible in a 2-10% of common cold in immunocompetent individuals (i.e. 229E, OC43E, NL63 and HKU1 types). However, other types can cause severe respiratory syndromes, such as SARS-CoV (Severe Acute Respiratory Syndrome Coronavirus) that caused an epidemics in 2002-2003, with origin on Guangdong (China) [Vijayanand et al 2004] and MERS-CoV (Middle Eastern Respiratory Syndrome Coronavirus) expanded in 2014-2015 through Saudi Arabia into Egypt, Oman and Qatar, from bats to dromedary camels, as the source of infection in humans [Aleanizy et al 2017].

At the end of 2019, a new SARS outbreak appeared in Wuhan (China) seafood market as the first focus of infection, becoming a pandemics in 2020 and spreading mainly into Europe and Asia, declaring general state of alarm in several countries, as Spain and Italy [Liao et al 2020, Zu et al 2020, Giovanetti et al 2020].

Coronavirus infections have normally low case fatality rates, with symptoms more severe than common cold, affecting mainly respiratory tract (cilia epithelium of the trachea, nasal mucosa and alveolar cells of the lung). Although the virus family is well-known and symptoms are similar to other coronavirus (fever, pneumonia, small pleural effusions), this specific type of virus presents considerable differences, as a higher infection/transmission and mortality rate, being a challenge for disease protection, prevention, diagnostic methods, vaccines and treatments [Wu et al 2020, Zhu et al 2020].

Virus pharmacology is based on preventive actions (vector-based or RBD-based vaccines), diagnostic methods (real time RT-PCR or ELISA tests) and antiviral drugs. SARS-CoV and MERS-CoV epidemics have expanded the use of several drugs, specially virus cycle inhibitors against coronaviruses (fusion inhibitors, RdRp inhibitors or PL2pro/3CLpro protease inhibitors) [Li G and De Clercq 2019, Raoult et al 2020]

In order to develop these methods, virus replication cycle must be simulated through computerized tools, specially for this virus family, with a complex replication cycle. Coronaviruses synthesize in a first stage a viral RNA-dependent RNA polymerase (RdRp) and multiple proteases, transcribes several subgenomic mRNAs and translates them progressively into viral proteins through several ribosomal pathways: (-1) programmed frameshift, leaky scanning and internal ribosome entry site (IRES) [Plant et al 2005]. Some mRNAs include genes encoding large polypeptide chains, which are cleaved through 3CLpro and PL2pro proteases, producing non-structural proteins as enzymes for catalyzing assembly and packaging of new viruses [Sawicki et al 2007, Fehr and Perlman 2015, Oxford et al. 2016]

Coronavirus.pro (2019-nCoV) App is a C++ code which simulates Coronavirus replication cycle. This software identifies virus types in short times and provides FASTA files of virus proteins, a list of RNA sequences (regulatory, packaging, transcription and translation) and secondary structures (stem-loops, helixes, palindromes, mirrors), once the virus genome has been sequenced [Ramos-Pascual 2019]. The code is supported by other RNA analysis tools, such as Vienna RNA package and Varna software [Gruver et al 2008, Darty et al 2009]. These results are useful as a first step with other bioinformatics and pharmacological tools in order to develop diagnostic methods, new vaccines and antiviral drugs.

## 2. The Coronavirus: classification, structure, genome and virus cycle

### 2.1 Classification

Coronaviruses (CoVs) are part of the order Nidovirales, from the family Coronaviridae and formed by several subtypes: Alpha-, beta-, gamma- and delta-coronavirus. Virus 229E and OC43, the first of being isolated and responsible of common cold, belong to alpha-coronavirus group I, while SARS and MERS are beta-coronavirus.

### 2.2 Structure of the virion

Coronaviruses have diameters from 100 to 160 μm with very large heavily glycosylated spikes (S) of 200kDa and 20 μm, placed around virus membrane as a crown, hence their name, in a trimer configuration (fig. 1).

Viral RNA genome is encapsulated in a helicoidal nucleocapsid phosphoprotein (N), known also as ribonucleoprotein (RNP), and enveloped into a virus particle with different membrane (M) and envelope glycoproteins (E).

Some coronaviruses include also a hemagglutinin-acetyleserase glycoprotein (HE) in the outer membrane. HEs helps during attachment, destroying certain sialic acid receptors in host-cell surface. Not all strains demonstrated hemagglutination, as observed only in beta-coronaviruses subgroup 2a (HCoV-HKU1, MHV) and toroviruses (BToV) [Brian et al. 1995, de Groot 2006]

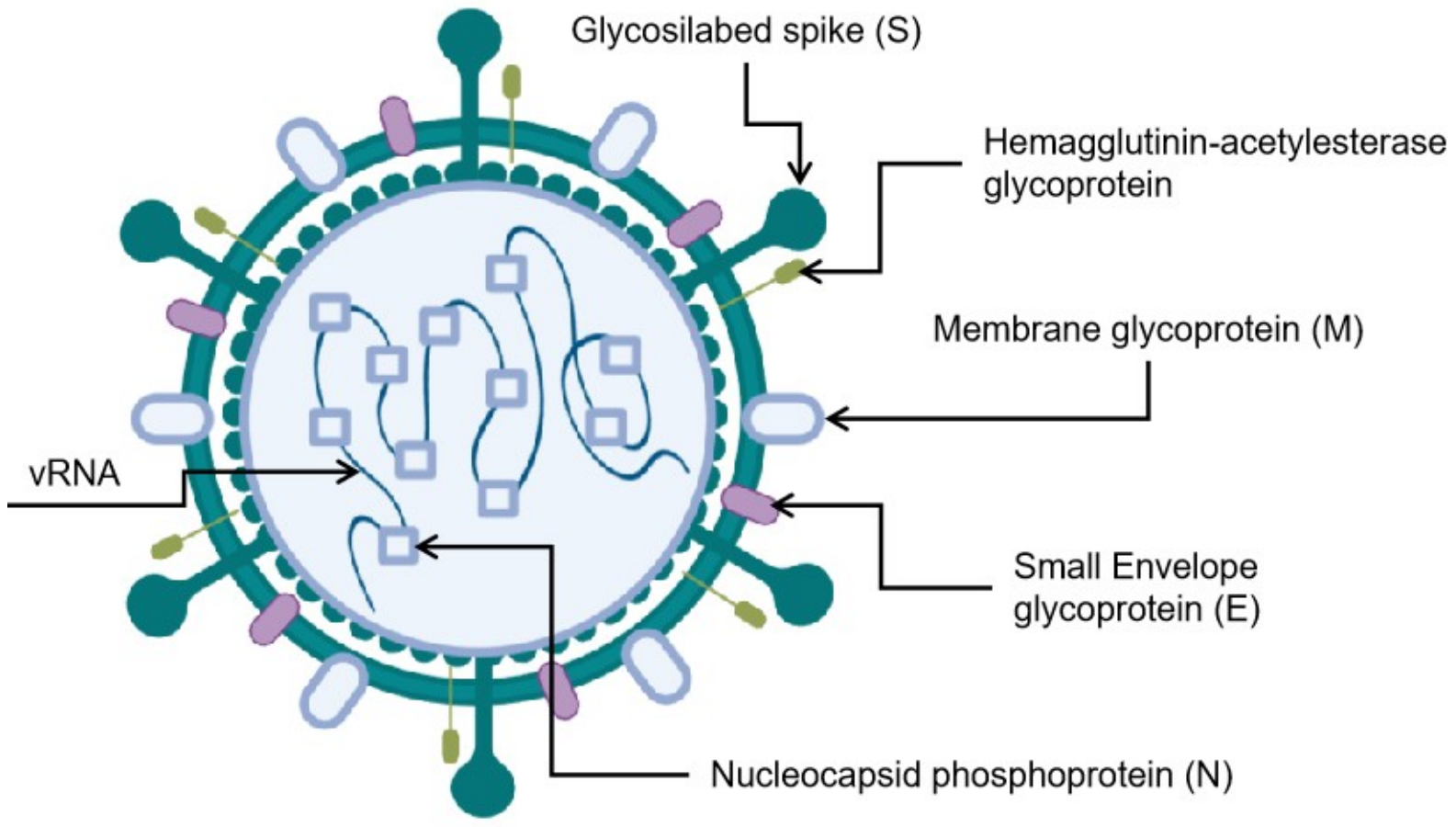

Fig. 1 - Structure of a general Coronavirus virion particle

### 2.3 Virus genome

Coronavirus genome is a type of positive single-stranded RNA of approximately 30kb, the largest of RNA viruses, 5'capped, 3'polyadenylated and infectious. This Poly(A) tail allows coronaviruses direct translation after infection without needing an intermediate transcription stage.

Transcription initiation is regulated in coronaviruses by several types of consensus transcription regulating sequences: TRS1-L, 5'-cuaaac-3', TRS2-L, 5'-acgaac-3' and merged into TRS3-L, 5'-cuaaacgaac-3'. These multiple TRS give place to several subgenomic polycistronic mRNAs, encoding structural, non-structural and accessory proteins. In case of some coronaviruses (i.e. MERS-CoV), transcription starts mainly in TRS2-L, while other coronaviruses start transcription indistinctly in all TRS, in a way of regulating protein frequencies [Sethna et al 1989, Irigoyen et al 2016].

Coronavirus genome includes multiple open-reading frames (ORF) containing genes which are transcribed by several transcription regulating sequences (TRS). Genes encoding non-structural proteins are placed at 5’ UTR (ORF1.1, ORF1.2 ...), whereas at 3’ UTR are genes for structural (N, M, E and S). These genes are interspaced with several accessory genes, encoding accessory proteins (AP), characteristics in number of each virus type. Some of these AP are not essential for in vitro or in vivo replication.

As transcription starts at different TRS in each subgenomic mRNA, the number of ORF genes is variable on virus type, and therefore the number of polypeptide chains. This produces different frequencies of non-structural proteins during virus cycle. For example, SARS produces ORF1.1 and ORF1.2 genes, whereas 2019-nCoV, produces ORF1.1 to ORF1.6 genes, synthesizing several groups of fusion proteins [van Boheemen et al. 2012]. A (-1) programmed slippery ribosome frameshift is placed approximately in the middle of ORF1 genes, then translated into polypeptides pp1a and pp1ab [Dinman 2012, Bock et al 2019]. Furthermore, Coronaviruses uses a leaky scanning mechanism (shunting) to synthesize proteins from overlapping ORF, translating different proteins from the same mRNA [Nakagawa et 2016].

Surface glycosyllabed Spike (S) is processed in some coronaviruses from a proteollytic cleavage of a spike precursor [Belouzard et al 2009]. The number of spike precursors is characteristic of each coronavirus. For example, in the case of SARS-CoV, two spike precursors (Sp1 and Sp2) are proteolytically cleaved, producing two surface glycosyllabed spikes (S1 an S2) and a protease fragment (S0).

Figure 2 shows a scheme of the main genes in Coronavirus family, including ORF1.1 with a -1 slippery ribosome frameshift. Genes transcribed from different TRS are placed in another line. Figure 3 presents a scheme of ORF1.1 gene and non-structural proteins (nsp1 to nsp16), including accessory protein AP2. Tables 1 and 2 summarize main genes and proteins of SARS-CoV and MERS-CoV coronaviruses.

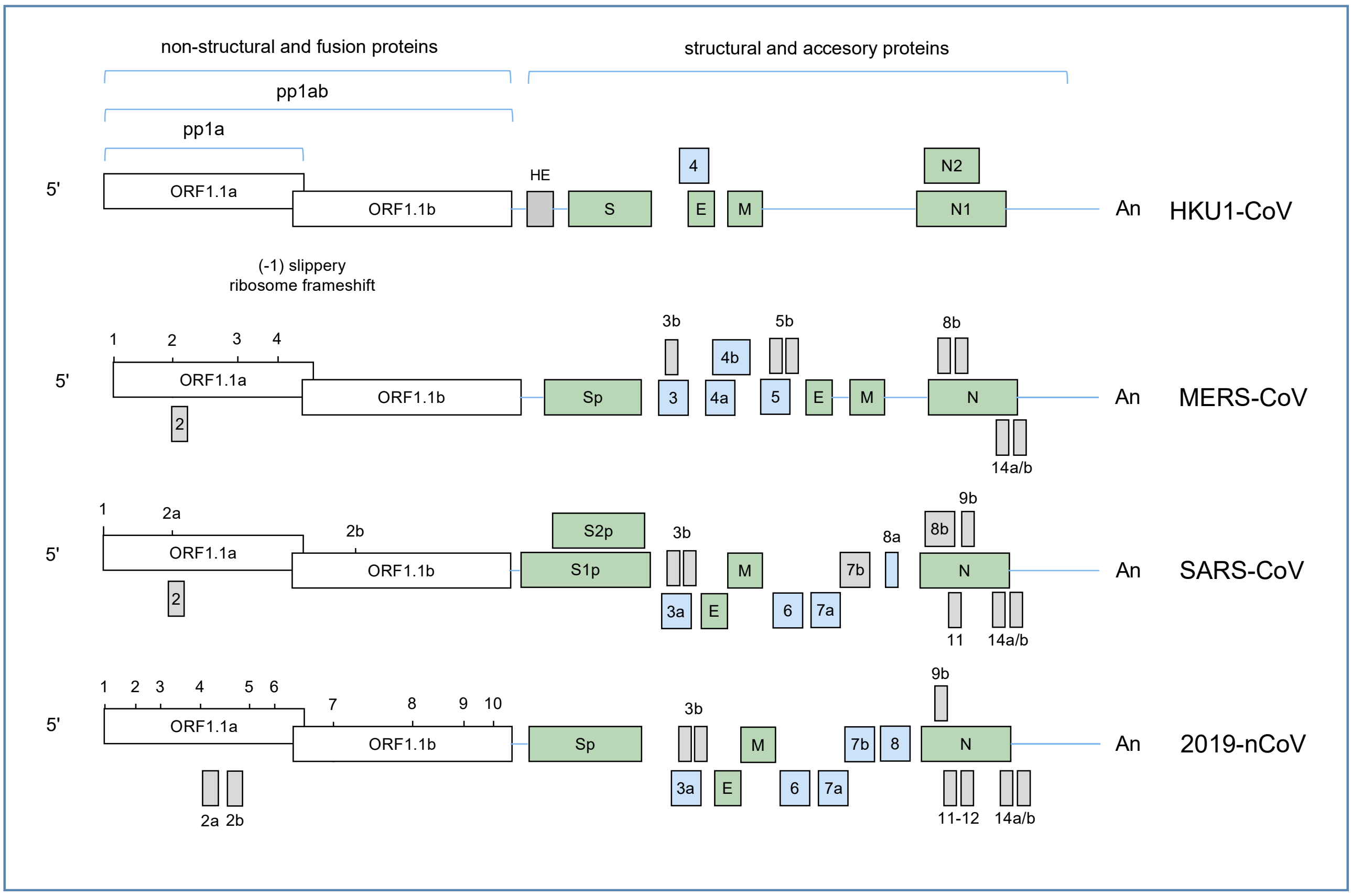

Fig. 2 – Scheme of the genes in viral genomes from Coronavirus family: non-structural proteins (white), structural (green), accessory (blue) and other ORF (grey). Fusion proteins and subgenomic mRNAs are not depicted.

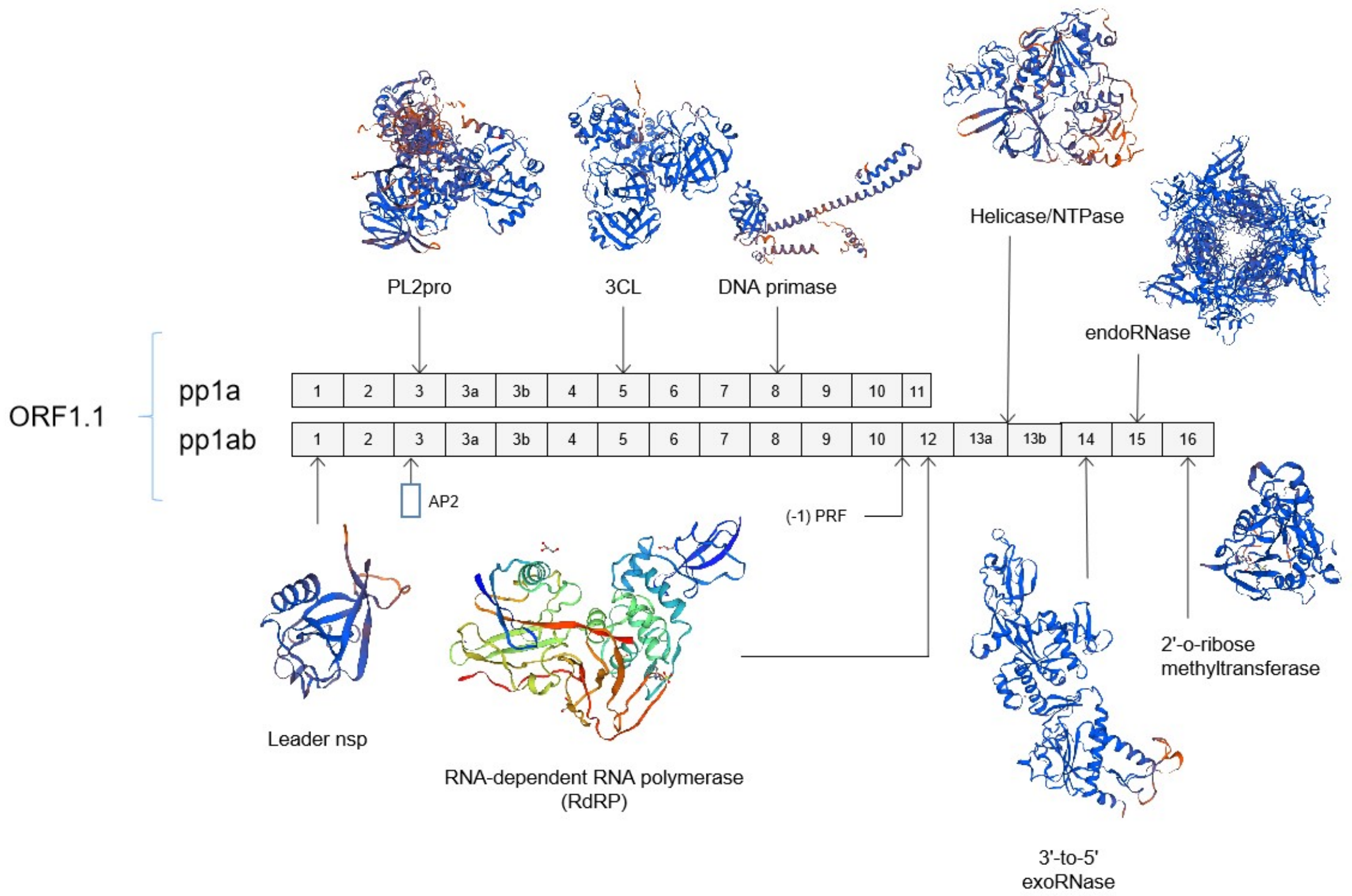

Fig. 3 – Scheme of the ORF1.1 gene and description of the non-structural proteins (nsp1 to nsp16) of SARS-CoV (SWISS Model) Accessory protein AP2 has been included

Table 1 - Summary of the main genes and characteristics of SARS-CoV [Xu et al 2003, Liu et al 2014] and MERS-CoV [Li et al 2019]

| Type | Coding Genes | Protein | Description |
|---|---|---|---|
| Non-structural | ORF 1.1 | pp1a | Polyprotein 1a |
| | | pp1ab | Polyprotein 1ab, [-1] PRF |
| Structural and accessory ORF | ORF 1.2 | AP2 | Accessory protein AP2, unknown |
| | | pp1a | Polyprotein 1a |
| | | pp1ab | Polyprotein 1ab, [-1] PRF |
| | ORF2 | Sp[(a)] | Surface Glycosylabed Spike precursor (Sp) |
| | | S | Surface Glycosylabed Spike (S) |
| | | S0 | Spike protease fragment (S0) |
| | ORF3a | AP3a | Viral pathogenesis, apoptosis induction, cell cycle arrest, modulation of NF-kb-mediated inflammation |
| | ORF3b | AP3b[(b)] | IRES translation, viral pathogenesis, not required for SARS-CoV replication |
| | ORF4 | E | Envelope membrane |
| | ORF5 | M | Transmembrane glycoprotein |
| | | AP5 | Unknown, only MERS |
| | ORF6 | AP6 | Type I IFN production and signaling inhibition, only SARS |
| | ORF7 | AP7a/b | Viral pathogenesis, apoptosis induction, cell cycle arrest, modulation of NF-kb-mmediated inflammation |
| | ORF8 | AP8a/b[(b)] | |
| | ORF9 | N | Nucleocapsid phosphoprotein |
| | ORF9b | AP9b[(b)] | Viral pathogenesis, apoptosis induction, cell cycle arrest, modulation of NF-kb-mmediated inflammation, named AP8b in MERS |
| | ORF11 | AP11[(b)] | Unknown, only SARS |
| | ORF14 | AP14a/b[(b)] | unknown |

(a) - Spike precursors length and number depends on virus type (Sp=S+S0) (b) TRS unknown

Table 2 - Description of non-structural proteins (Polyprotein pp1ab) of SARS and MERS coronavirus [Chen et al 2020]

| Protease | Protein | Comments |
|---|---|---|
| PL2pro | nsp1 | Leader protein, suppress antiviral host response, promotes degradation of host mRNAs, inhibiting IFN signaling |
| | nsp2 | unknown |
| | nsp3 | ADP-ribose 1-phosphatase, PL2pro (papain-like protease 2) |
| | nsp3a | unknown |
| | nsp3b | unknown |
| 3CLpro | nsp4 | DMV formation, complex with nsp3 |
| | nsp5 | 3C-like (3CLpro), Mpro, polypeptides cleaving |
| | nsp6 | Restricting autophagosome expansion, DMV formation |
| | nsp7 | Cofactor with nsp8 and nsp12 |
| | nsp8a | DNA primase, cofactor with nsp7 and nsp12 |
| | nsp8b | |
| | nsp9 | Dimerization and RNA/DNA binding activity |
| | nsp10 | interacts with nsp14 and nsp16 [Bouvet et al 2010,2012] |
| | nsp11 | Short peptide at pp1a end |
| | nsp12 | RNA-dependent RNA polymerase (RdRp) |
| | nsp13a | Helicase, NTPase nucleoside 5’ triphosphatase (ZD, NTPase/HEL) |
| | nsp13b | |
| | nsp14 | 3’-to-5’ exoribonuclease (nuclease ExoN homolog) |
| | nsp15 | Endoribonuclease (endoRNAse), evasion of dsRNA sensors |
| | nsp16 | S-adenosylmethionine-dependent ribose 2'-O- methyltransferase (2'-O-MT) |

2.4 Virus replication cycle

As other viruses, Coronavirus employs glycosillabed spikes placed in the outer surface to attach specific receptors of host cells (i.e. APN/ACE2/DPP4 receptor). Specially, betacoronaviruses attach to angiotensin-converting enzyme 2 (ACE2) receptor, a membrane protein expressed mainly in the surface of epithelial cells of the pulmonary alveolus [Jia et al 2005]. Once attachment is carried out, viral transmembrane is fused through an endocytotic pathway and viral RNA is released into the cell cytoplasm [Wang et al 2008]. Viral RNA genome is positive stranded, 5' capped and 3'polyadenylated, therefore it can be directly translated into proteins by host-cell ribosomes. Specifically, ORF1 gene contains several non-structural proteins in a polypeptide complex, that once translated, is are catalytically autoprocessed by 3CL/2PL proteases, and assembled into a replicase-transcriptase complex with RNA-dependent RNA activity (RdRp). At this point, several subgenomic mRNAs are produced by transcription and translated into structural (N, M, E and S), non-structural and accessory proteins (AP), assembling new virions. These virus particles are formed on smooth-walled vesicles located between the ER and the Golgi, named as ERGIC (Endoplasmic Reticulum Golgi Intermediate Compartment). Once these vesicles fuse with the plasma outer membrane, virions are released to continue infection (see fig 4).

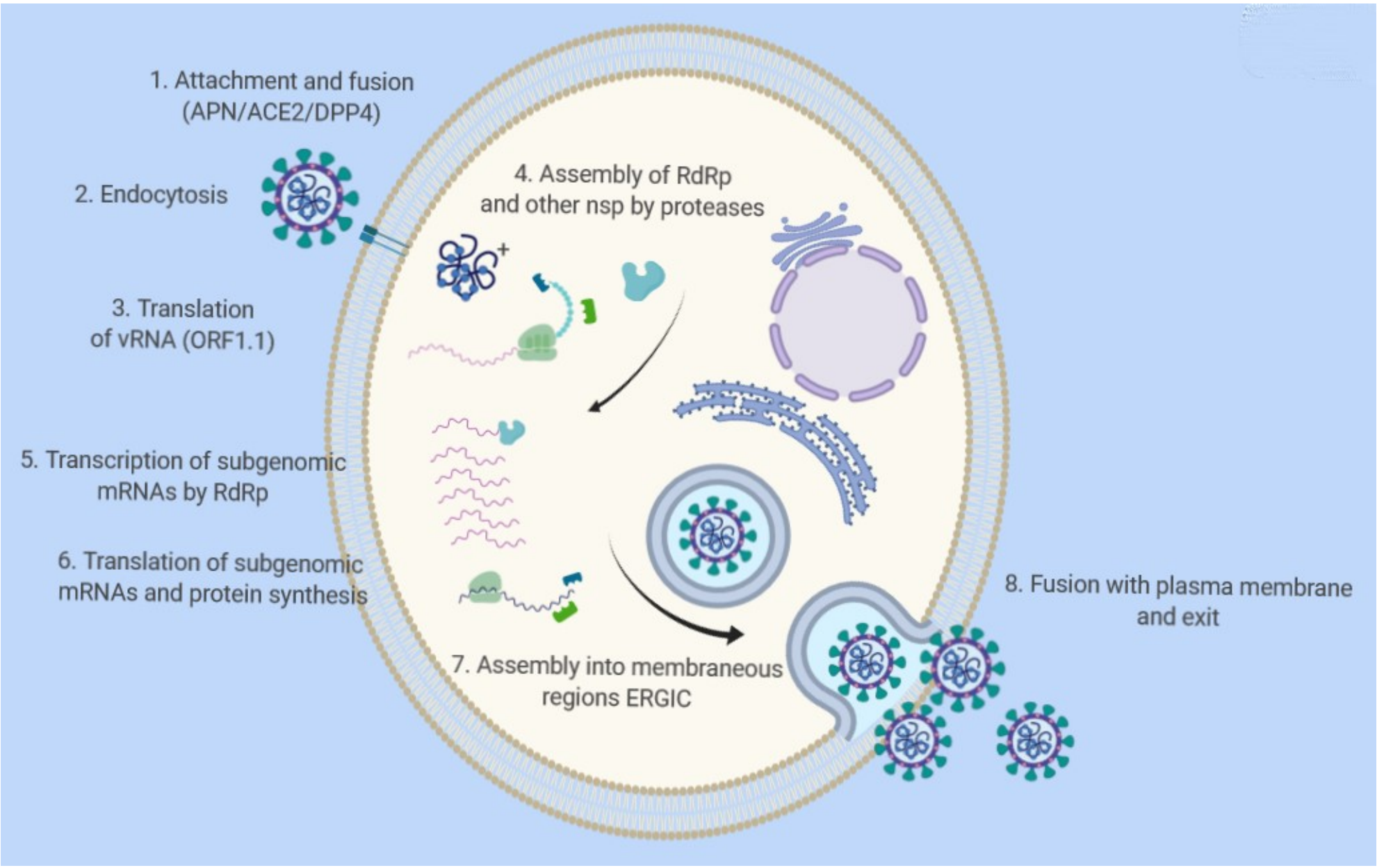

Figure 4 – Scheme of a Coronavirus cycle replication: [1] Attachment and fusion (APN/ACE2/DPP4), [2] endocytosis, [3] translation of vRNA (ORF1.1), [4] assembly of RNA-dependent RNA polymerase (RdRp) and other non-structural proteins (nsp) by proteases, [5] transcription of subgenomic mRNAs by RdRp, [6] translation of subgenomic mRNAs and protein synthesis, [7] assembly into membraneous regions ERGIC and [7] fusion with plasma membrane and exit (BioRender http://app.biorender.io)

## 3. Coronavirus.pro

### 3.1 Description

Coronavirus.pro is a module of Virus.pro, a C++ software application developed in modules that simulate mainly RNA and DNA virus replication cycles: Ebola, HIV-1, HCV (Hepatitis C), CoV, HSV1 (Human Herpes Virus), PV1 (Poliovirus 1, Mahoney)  The software reproduces several virus cycle replication stages, from attachment and fusion to virion exit from host-cell, focusing into more complex stages, such as subgenomic mRNAs translation, protein synthesis and protease catalytic processing (see fig. 5 and 6).

Virus.pro contains a set of RNA/DNA databases and protein databases to scan viral genome and protein sequences for recognized motifs, reconstruct secondary structures (helixes, stem-loops, palindromes, mirrors) and identify RNA-protein interaction regions. The software is supported with other applications, as Vienna RNA package, for bracket-dot notation and Varna for plotting (see fig. 7) [Gruber et al 2008, Darty et al 2009]. The code contains also machine-learning algorithms, in which new virus, RNA/DNA sequences and proteins can be included to the internal databases to future identifications and analysis. The software has been validated with other bio-informatic tools, as Blastp or Swiss-Model [Altschul et al 1990, Camacho et al 2008, Waterhouse et al 2018, Ramos-Pascual 2019].

```
|====================================================================|
|     Virus.pro                                                      |
|====================================================================|
|     Analysis of virus cycle replication:                           |
|                                                                    |
|           [1] CoV (SARS, MERS)                                     |
|           [2] Ebola                                                |
|           [3] HIV-1                                                |
|           [4] HCV (Hepatitis C virus)                              |
|           [5] HSV1 (Human Herpes Simplex virus 1)                  |
|           [6] PV1 (Poliovirus 1, Mahoney)                          |
|           [7] Other viruses                                        |
|                                                                    |
|     Analysis of RNA/DNA or protein sequences:                      |
|                                                                    |
|           [8]  Sequence analysis                                   |
|                                                                    |
|     RNA / protein interactions:                                    |
|                                                                    |
|           [9]  RNA-protein interaction                             |
|           [10] RNA-RNA interaction                                 |
|           [11] Protein-protein interaction                         |
|                                                                    |
|====================================================================|
```

Fig. 5 – Software Virus.pro for simulating RNA/DNA virus replication cycles

```
|====================================================================|
|                                                                    |
|   CoronaVirus.pro - SARS/MERS/2019-nCoV Replication Cycle Simulation   |
|                                                                    |
|====================================================================|
|                                                                    |
|   (1)  Attachment and fusion (APN/ACE2/DPP4 receptor)              |
|   (2)  Endocytosis                                                 |
|   (3)  Translation of vRNA (ORF1.1)                                |
|   (4)  Assembly of RNA-dependent RNA polymerase (RdRp) and         |
|           other non-structural proteins (nsp) by proteases         |
|   (5)  Transcription of subgenomic mRNAs by RdRp                   |
|   (6)  Translation of subgenomic mRNAs and protein synthesis       |
|   (7)  Assembly into membraneous regions ERGIC                     |
|   (8)  Fusion with plasma membrane and exit                        |
|                                                                    |
|====================================================================|
```

Fig. 6 – Software Coronavirus.pro for simulating SARS/MERS/2019-nCoV virus replication cycle

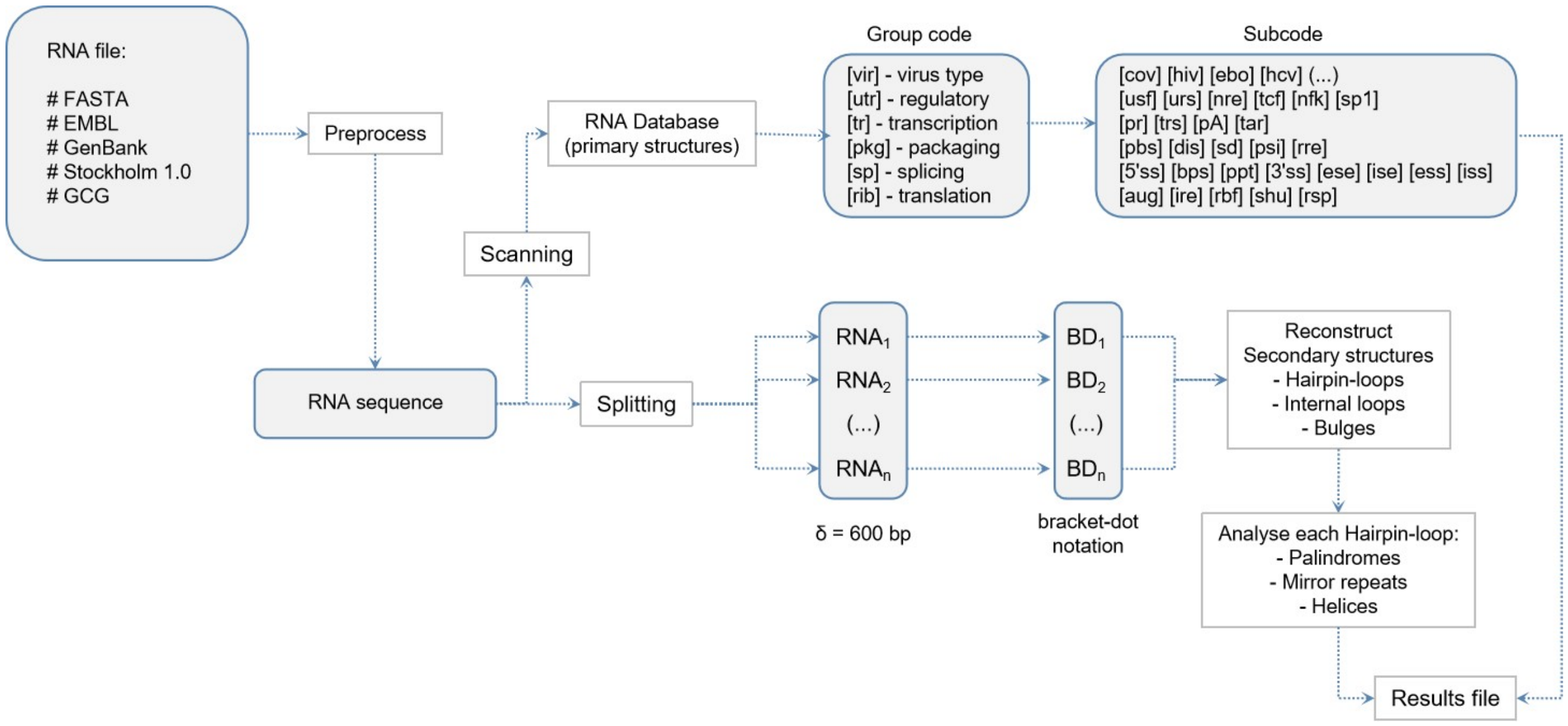

Fig. 7 - Scheme of the Virus.pro software (RNA module)

Coronavirus.pro includes a preprocessor to convert viral genome sequence file into a plain sequence format (nucleotides list). Preprocessor supports genomes in formats: FASTA, EMBL, GenBank, Stockholm 1.0 and GCG (see fig 8).

```
!====================================================================================================================
!  Software for pre-process virus genome file into plain sequence format (nucleotides):
!    (RNA/DNA virus genome is saved with output name: file00.tmp)
!     - Enter  filename with extension (i.e. hsv.fasta)
!     - Select filetype: (a) FASTA (b) EMBL (c) GenBank (d) Stockholm 1.0 (e) GCG (f) Others
!====================================================================================================================
```

Fig. 8 – Software for preprocessing viral genome sequence file. Supported formats: FASTA, EMBL, GenBank, Stockholm 1.0 and GCG

## 3.2 Simulation of translation, protein synthesis and protease cleavage sites

Each subgenomic mRNA transcribed by coronavirus RNA-dependent RNA polymerase (RdRp) is translated into virus proteins by host-cell ribosomes. ORF1.1 and ORF1.2 in SARS-CoV, ORF1.1 to ORF1.4 in MERS-CoV and ORF1.1 to ORF1.6 in 2019-nCoV encode a polypeptide chain of variable length, with a -1 programmed ribosomal frameshift (PRF). This frameshift is followed by a pseudoknot structure located few nucleotides downstream and is able to change reading frame -1 position backward, translating an alternative polypeptide (pp1ab with frameshift). Frameshifting probability is approximately between 5-10%. In same cases, translation starts after a leaky scanning (shunting) [Dinman 2012, Bock et al 2019] or Internal Ribosomal Entry Site (IRES) [Bonnal et al 2003, Mokrejs et al 2006]

After this, viral proteases cleave with enzymatic activity these polypeptides at specific cleavage sites to synthesize non-structural proteins [Kiemer et al 2004]. Protease cleavage sites are predicted depending on protease family (aspartic, cysteine, metallo or serine protease) and through specific cleavage patterns [Song et al. 2012] Coronavirus proteases are a papain-like ($PL2^{pro}$) and a cysteine 3C-like proteinase ($3CL^{pro}$) synthesized from nsp3 and nsp5, respectively [Chen et al 2005].

Coronavirus.pro simulates proteolytic effect of coronavirus proteases PL2pro and 3CLpro though multiple protease pattern sequences (table 3). Most of these sequences have been previously validated in some research studies and others have been proposed by comparison with protein databases, as UniProt or NCBI, and recursive simulations with the software [Kiemer et al 2004, Sulea et al 2006, Ramos-Pascual 2019].

Table 3 - Protease cleavage site sequences for Coronavirus proteins (SARS -CoV and MERS-CoV)

| PCS | Sequence | SARS-CoV | MERS-CoV |
|---|---|---|---|
| [1.1] | VSQIQ↓ SRLT | S1/S2-S0 | - |
| [1.2] | GKIQD↓SLSST | S1/S2-S0 | - |
| [1.3] | GAMQT↓GFTTT | - | S1/S2-S0 |
| [2] | YPKLQ↓ASQAW | M1-M2 | - |
| [3] | SNNLQ↓GLEN | - | N1-N2 |
| [4] | ETRVQ↓CSTN | - | N2-N3 |
| [5] | ELNGG↓AVTRY | nsp1-nsp2 | - |
| [6] | DPKGK↓YAQNL | | nsp1-nsp2 |
| [7] | RLKGG↓APIKG | nsp2-nsp3 | nsp2-nsp3 |
| [8] | KSSVQ↓ SVAG | nsp3-nsp3a | - |
| [9] | KNTVK↓SVGKF | nsp3-nsp3a | - |
| [10] | AQGLK↓KFYKE | - | nsp3-nsp3a |
| [11/4] | ETRVQ↓CSTN | nsp3a-nsp3b | nsp3a-nsp3b |
| [12] | SLKGG↓KIVST | nsp3b-nsp4 | - |
| [13] | KIVGG↓APTWF | - | nsp3b-nsp4 |
| [14] | SAVLQ↓SGFRK | nsp4-nsp5 | nsp4-nsp5 |
| [15] | GVTFQ↓GKFK | nsp5-nsp6 | - |
| [16] | GVVMQ↓SGVRK | - | nsp5-nsp6 |
| [17] | VATVQ↓SKMSD | nsp6-nsp7 | - |
| [18] | VATLQ↓AENV | nsp7-nsp8a | - |
| [19] | VAAMQ↓SKLTD | nsp8a-nsp8b | nsp6-nsp7 |
| [20] | HSVLQ↓APMST | - | nsp7-nsp8a |
| [21] | AVKLQ↓NNELS | nsp8b-nsp9 | nsp8a-nsp9 |
| [22] | TVRLQ↓AGNAT | nsp9-nsp10 | nsp9-nsp10 |
| [23] | EPLMQ↓ SADA | nsp10-nsp11/nsp12 | - |
| [24] | ALPQS↓KDSNF | - | nsp10-nsp11/nsp12 |
| [25] | HTVLQ↓AVGAC | nsp12-nsp13a | nsp12-nsp13a |
| [26/18] | VATLQ↓AENV | nsp13a-nsp14 | nsp13a-nsp13b |
| [27] | YKLQS↓QIVTG | - | nsp13b-nsp14 |
| [28] | FTRLQ↓SLENV | nsp14-nsp15 | - |
| [29] | TKVQG↓LENIA | - | nsp14-nsp15 |
| [30/2] | YPKLQ↓ASQAW | nsp15-nsp16 | nsp15-nsp16 |

Cleavage sites are identified with a coarse approximation in which each protease cleavage sequence (A) scans through each protein aminoacid sequence (B). If ka and kb are respectively the amino acid length of A sequence and protein B, protein is cleaved at positions with the highest Levenshtein distance, calculated as:

```
for (i = 1; i <= ka; i++) d[i][0] = i; for (i = 1; i <= kb; i++) d[0][i] = i;

for (i = 1; i <= ka; i++)

        for (j = 1; j <= kb; j++)

                c = 0;

                if (a[i - 1] == b[j - 1]) { c = 0; }

                else { c = 1; };

                d[i][j] = min(d[i - 1][j] + 1, d[i][j - 1] + 1, d[i - 1][j - 1] + c)

        }

}
```

## 4. Results and discussion

### 4.1 Virus identification: comparison with SARS-CoV and MERS-CoV

Coronavirus.pro has been used with sequence MN908947 (Wuhan seafood market pneumonia virus isolate Wuhan-Hu-1, complete genome) This sequence is specific of the virus type that caused the outbreak in Wuhan of Hubei province (China), the first infection focus (23-jan-2020), which has been named as 2019-nCoV [NCBI database]

The code has been also applied to other coronavirus types, such as sequences NC004718 (SARS coronavirus, complete genome) and NC019843 (MERS Middle East respiratory syndrome coronavirus, complete genome) [Snijder et al 2003, Moreno et al 2017], which have been taken as reference sequences. Other sequences have been also applied to compare similarity with 2019-nCoV virus genome (see table 4)

Table 4 - Summary of Coronavirus sequence files applied to Coronavirus.pro software

| Virus | Sequence | Date | Description | bp | Comments |
|---|---|---|---|---|---|
| 2019-nCoV | MN908947 | 23-JAN-2020 | Wuhan seafood market pneumonia virus isolate Wuhan-Hu-1 (2019-nCoV) | 29903 | [ref] |
| MERS-CoV | NC019843 | 13-AUG-2018 | Middle East respiratory syndrome coronavirus | 30119 | [ref] |
| SARS-CoV | NC004718 | 13-AUG-2018 | SARS-CoV coronavirus | 29751 | [ref] |
| | KY417149 | 18-DEC-2017 | Bat SARS-like coronavirus isolate Rs4255 | 29743 | - |
| | AY278488 | 01-SEP-2009 | SARS coronavirus BJ01 isolate genome sequence | 29725 | - |

### 4.2 Coronavirus RNA structure

#### 4.2.1 Regulatory regions: 5’utr

Beta-coronaviruses have several stem-loop structures in the 5’utr region (SL1 toSL5C). The first transcription regulating sequence (TRS-L) is placed around the same positions in all beta-coronaviruses (SL3).

MERS-CoV presents a 5’utr region of 356bp, with a TRS-L without SL3. Stem-loop SL4b is only present in 2019-nCoV and MERS-CoV. Furthermore, there are several regulatory sequences (RS) with unknown functionality. Stem-loops SL6 and SL7 are placed into the adjacent ORF1 coding region. [Yang and Leibowitz 2015, Madhugiri et al 2018]

SARS-CoV and 2019-nCoV have a common 5’ utr region of approximately 300bp, with several negative regulatory elements: NRE-I (IL-2R/EBS), NRE-II (Ap1) and NRE-III (Ap1) at positions 101/153/250 and 104/154/251, respectively. Stem-loop SL4b is absent in SARS-CoV, whereas in 2019-nCoV contains a palindrome sequence (-UAAUUA//UAAUUA-) with an unknown function.

Figures 9 to 11 shows 5’utr regions in MERS-CoV, SARS-CoV and 2019-nCoV coronaviruses, obtained through bracket-dot notation from Vienna RNA package and plotted with Varna software.

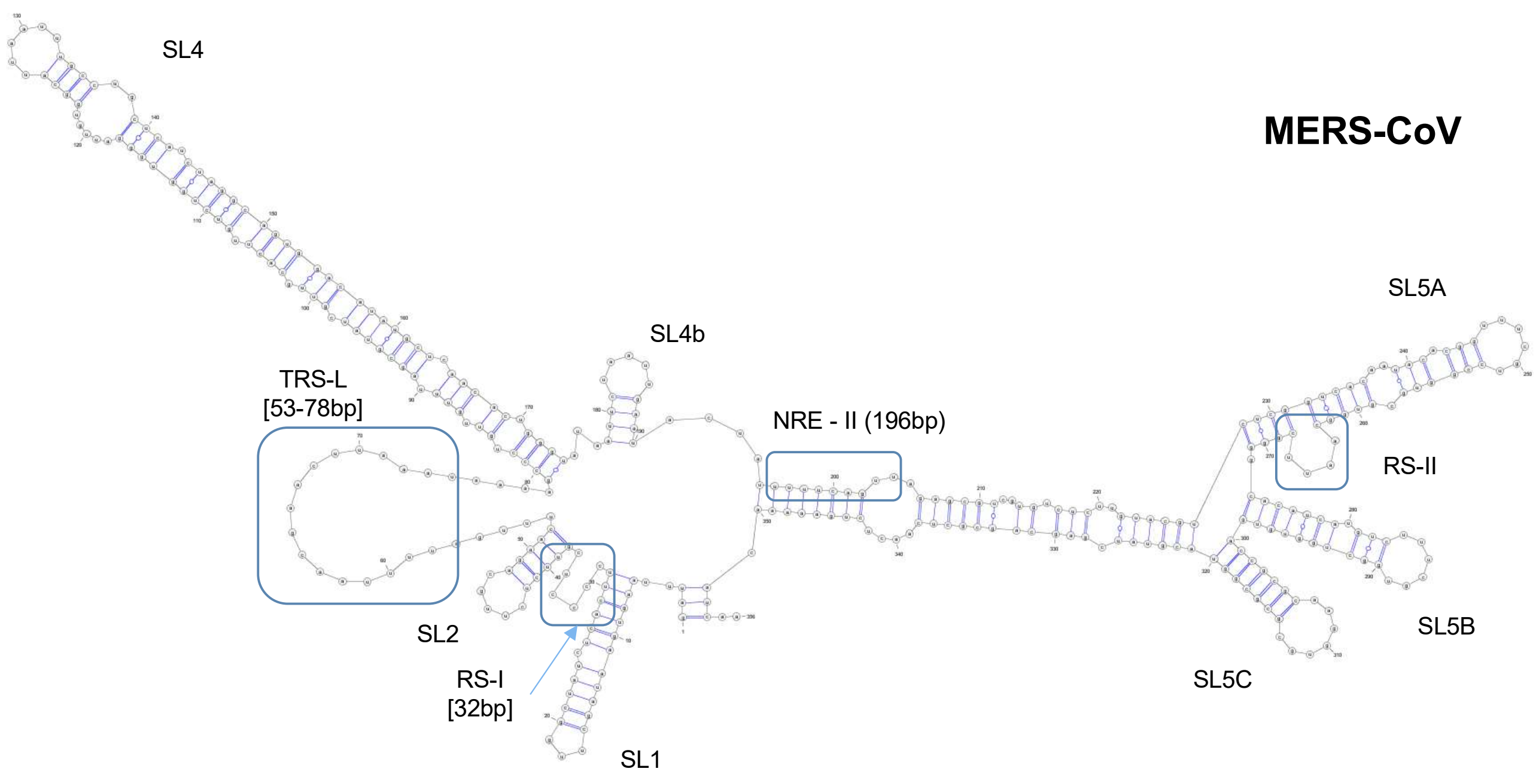

Fig. 9 – MERS-CoV [1-356bp] - Scheme of the 5’ utr secondary structures (SL1-SL5C)

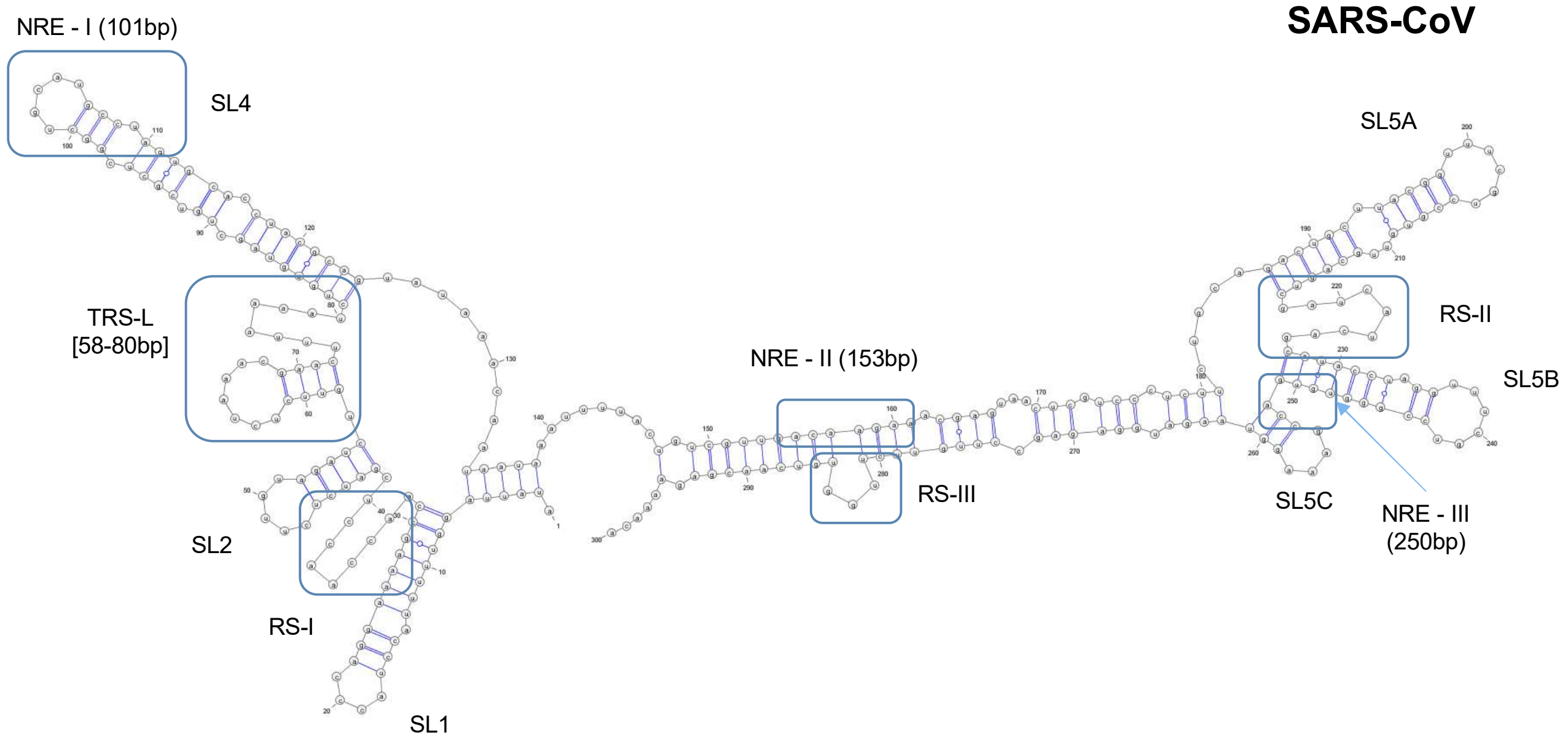

Fig. 10 – SARS-CoV [1-300bp] - Scheme of the 5’ utr secondary structures (SL1-SL5C)

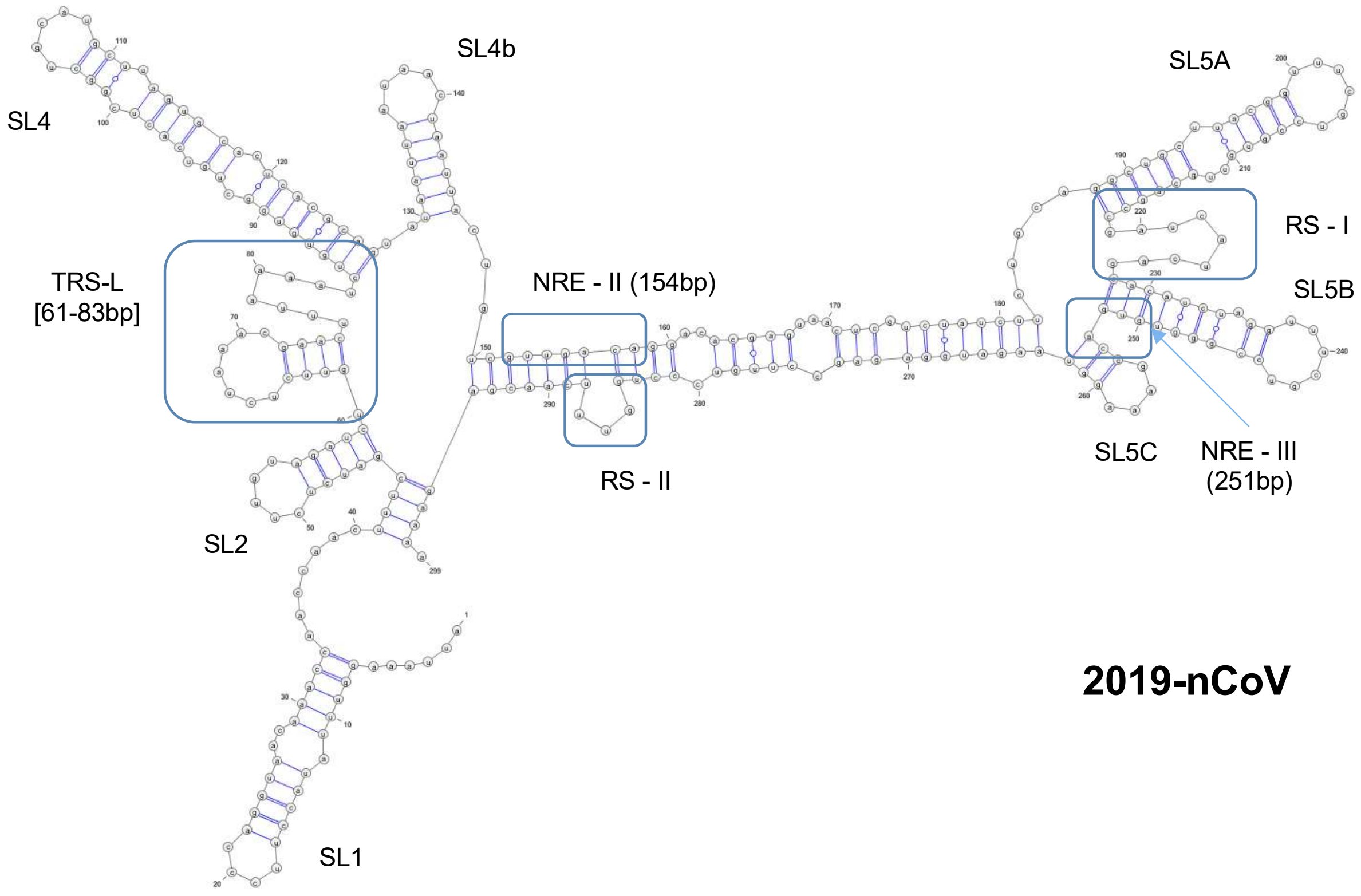

Fig. 11 – 2019-nCoV [1-300bp] - Scheme of the 5’ utr secondary structures (SL1-SL5C)

4.2.2 Regulatory regions: 3’utr

Beta-coronaviruses have a short 3’utr region of approximately 100bp with a stem-loop of variable length followed with a poly(A) tail. This stem-loop, in the case of MERS-CoV is 27bp, including other recognition sequences, different as SARS-CoV and 2019-nCoV coronaviruses, as presented in figure 12. The 3’ utr region of these coronaviruses contains also a conserved pseudo-knot structure of approximately 55bp [Lin et al 1996, Yang and Leibowitz 2015, Peng et al 2016].

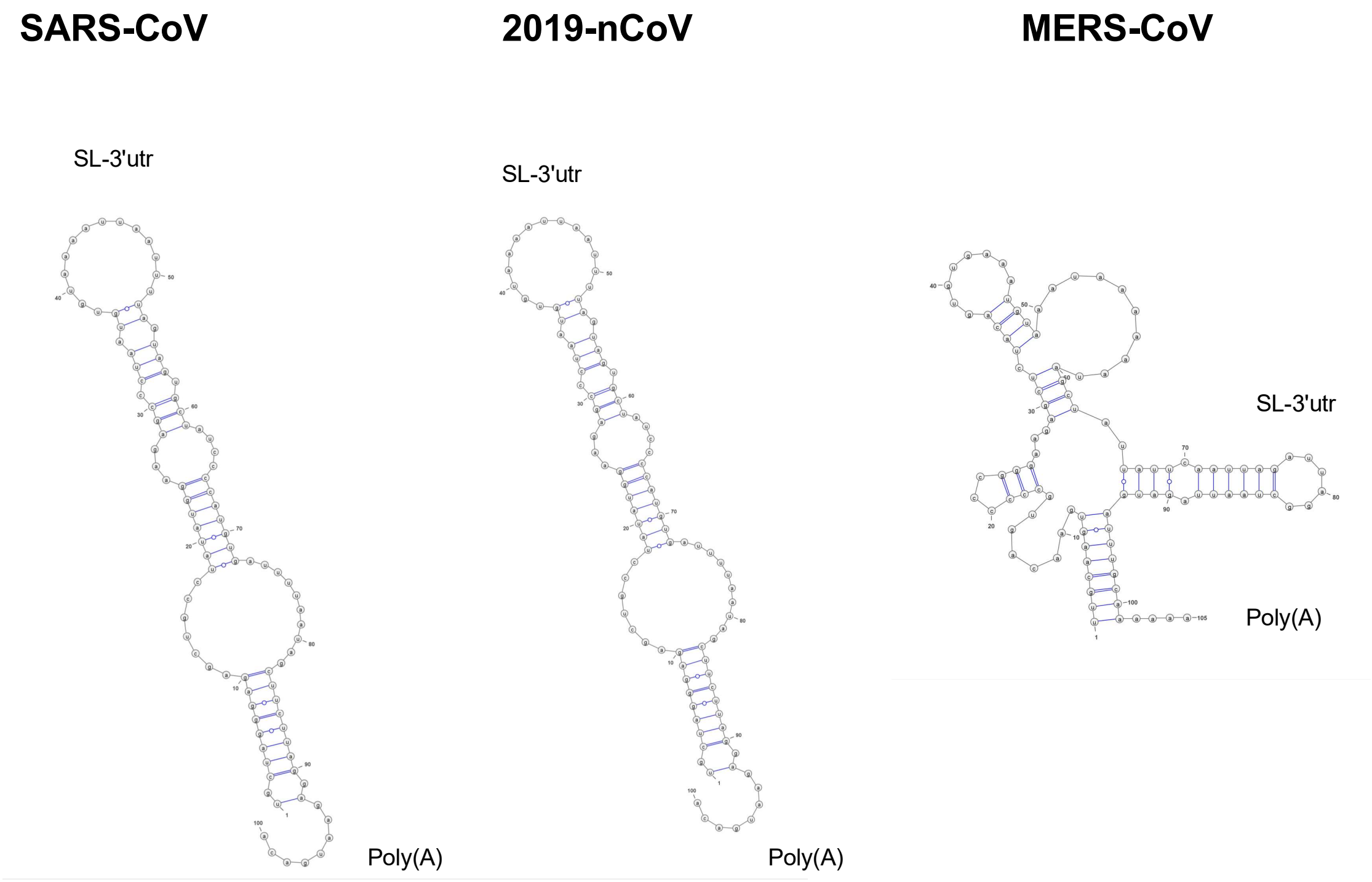

Fig. 12 – Scheme of the 3’ utr secondary structures [1-100bp]: stem-loop (SL) and Poly(A) tail

## 4.3 Protease cleavage sites

Coronavirus.pro has predicted all canonical protease cleavage sites in coronavirus polypeptides, from nsp1 to nsp16. Furthermore, the software has identified a protease cleavage site in spike glycoprotein precursors (Sp) in all coronaviruses. This cleavage site splits spike precursors (Sp) into S/S0 proteins, where S0 is the same protease fragment in both precursors.

Another cleavage site has been predicted in membrane protein (M) of 2019-nCoV virus, producing fragments M1/M2 In the case of MERS-CoV nucleocapsid (N), the software has found two protease cleavage sites, N1/N2 and N2/N3. In addition to these cleavage sites, the code has identified other protease cleavage sites giving place to some hypothetical proteins, as nsp3a↓nsp3b, nsp8a↓nsp8b and nsp13a↓nsp13b (see table 5). These cleavage sites must be discussed in detail and supported with other methods.

Table 5 - Protease cleavage sequences predicted by Coronavirus.pro

| | Protein | | PCS | SARS-CoV | PCS | 2019-nCoV | PCS | MERS-CoV |
|---|---|---|---|---|---|---|---|---|
| Structural proteins | S1 & S2 | S0 | [1.1] | SQIQE↓SLTTT | [1.2] | GKIQD↓SLSST | [1.3] | GAMQT↓GFTTT |
| | M1 | M2 | [2] | YKLGA↓SQRVG | [2] | YKLGA↓SQRVA | - | - |
| | N1 | N2 | - | - | - | - | [3] | NRLQA↓LESGK |
| | N2 | N3 | - | - | - | - | [4] | QRVQG↓SITQR |
| Non-structural proteins | nsp1 | nsp2 | [5] | ELNGG↓AVTRY | [5] | ELNGG↓AYTRY | [6] | DPKGK↓YAQNL |
| | nsp2 | nsp3 | [7] | RLKGG↓APIKG | [7] | LKGGA↓PTKVT | [7] | RLKGG↓APVKK |
| | nsp3 | nsp3a | [8] | NSVKS↓VAKLC | [9] | KNTVK↓SVGKF | [10] | AQGLK↓KFYKE |
| | nsp3a | nsp3b | [11/4] | TRVEC↓TTIVN | [11/4] | TRVEC↓TTIVN | [11/4] | TRVEA↓STVVC |
| | nsp3b | nsp4 | [12] | SLKGG↓KIVST | [12] | ALKGG↓KIVNN | [13] | KIVGG↓APTWF |
| | nsp4 | nsp5 | [14] | SAVLQ↓SGFRK | [14] | SAVLQ↓SGFRK | [14] | GVLQS↓GLVKM |
| | nsp5 | nsp6 | [15] | GVTFQ↓GKFKK | [15] | VTFQS↓AVKRT | [16] | GVVMQ↓SGVRK |
| | nsp6 | nsp7 | [17] | VATVQ↓SKMSD | [17] | VATVQ↓SKMSD | [19] | VAAMQ↓SKLTD |
| | nsp7 | nsp8a | [18] | ATLQA↓IASEF | [18] | ATLQA↓IASEF | [20] | SVLQA↓TLSEF |
| | nsp8a | nsp8b | [19] | AAMQR↓KLEKM | [19] | AAMQR↓KLEKM | [21] | AVKLQ↓NNEIK |
| | nsp8b | nsp9 | [21] | AVKLQ↓NNELS | [21] | AVKLQ↓NNELS | | |
| | nsp9 | nsp10 | [22] | TVRLQ↓AGNAT | [22] | TVRLQ↓AGNAT | [22] | TVRLQ↓AGSNT |
| | nsp10 | nsp11 & nsp12 | [23] | EPLMQ↓SADAS | [23] | PMLQS↓ADAQS | [24] | ALPQS↓KDSNF |
| | nsp12 | nsp13a | [25] | HTVLQ↓AVGAC | [25] | HTVLQ↓AVGAC | [25] | TTLQA↓VGSCV |
| | nsp13a | nsp13b | [26/18] | VATLQ↓AENVT | [26/18] | VATLQ↓AENVT | [26/18] | ATLTA↓PTIVN |
| | nsp13b | nsp14 | | | | | [27] | YKLQS↓QIVTG |
| | nsp14 | nsp15 | [28] | FTRLQ↓SLENV | [28] | FTRLQ↓SLENV | [29] | TKVQG↓LENIA |
| | nsp15 | nsp16 | [30/2] | YPKLQ↓ASQAW | [30/2] | YPKLQ↓SSQAW | [30/2] | TFYPR↓LQASA |

## 4.4 Subgenomic mRNA

Most frequent transcription regulating sequence (TRS) in MERS-CoV is TRS2 (5'-acgaac-3'). MERS-CoV transcribes 11 subgenomic mRNAs, with several ORFs translating polypeptide chains (ORF1.1 to ORF1.4) and several fusion proteins [Li et al 2019].

ORF3b is translated with an Internal Ribosome Entry Site (IRES). Other ORF encode some proteins (AP8b, AP7b, AP9b and AP14) with unknown transcription regulating sequences and functionality [Narayanan et al 2008]. ORF8b overlaps nucleocapside gene (N) and encodes an accessory protein AP8b, called AP9b in SARS. Although it is not translated directly by any TRS, some studies have found antibodies specific to this protein in both in vitro and in vivo samples [Sharma et al 2011]

Tables 6 and 7 present a summary of ORF and proteins identified with Coronavirus.pro in MERS-CoV.

Table 6 - MERS-CoV open-reading frames (ORF) and proteins identified with Coronavirus.pro

| mRNA-TRS[(a)] | j (bp) | ORF | proteins (Aa) | | Fusion protein (Aa) | | Comments |
|---|---|---|---|---|---|---|---|
| 1-[2] | 63 | ORF1.1 | pp1a | 4391 | - | - | [nsp1-nsp11] |
| | | | pp1ab | 7078 | - | - | [nsp1-nsp10], [nsp12-nsp16] |
| 2-[1] | 3904 | ORF1.2 | AP2 | 58 | - | - | unknown |
| | | | pp2a | 3022 | pp2/nsp3 | 846 | [nsp3a-nsp11] |
| | | | pp2ab | 5709 | pp2/nsp3 | 846 | [nsp3a-nsp10], [nsp12-nsp16] |
| 3-[1] | 11815 | ORF1.3 | pp3a | 487 | pp3/nsp7 | 24 | [nsp8-nsp11] |
| | | | pp3ab | 3174 | pp3/nsp7 | 24 | [nsp8-nsp10], [nsp12-nsp16] |
| 4-[2] | 12751 | ORF1.4 | pp4a | 226 | pp4/nsp9 | 72 | [nsp10-nsp11] |
| | | | pp4ab | 2913 | pp4/nsp9 | 72 | [nsp10-nsp16] |
| 5-[2] | 21405 | ORF2 | Sp | 1354 | - | - | Surface glycoprotein spike precursor |
| | | | S | 1010 | - | - | Surface glycoprotein spike |
| | | | S0 | 344 | - | - | S0 protease fragment |
| 6-[2] | 25521 | ORF3 | AP3 | 103 | - | - | Accessory protein AP3 |
| 7-[1] | 25843 | ORF4a | AP4a | 109 | - | - | Accessory protein AP4a |
| 8-[2] | 25928 | ORF4b | AP4b | 246 | - | - | Accessory protein AP4b |
| 9-[2] | 26833 | ORF5 | AP5 | 224 | - | - | Accessory protein AP5 |
| 10-[2] | 27583 | ORF6 | E | 82 | - | - | Envelope protein |
| 11-[2] | 27838 | ORF7 | M | 219 | - | - | Membrane protein |
| | | ORF8a | N | 413 | - | - | Nucleocapsid phosphoprotein |
| | | | N1/N2/N3 | 223/166/25 | - | - | N1/N2/N3 protease fragments |

(a) Transcription regulating sequence: [1] TRS1 - 5'-cuaaac-3' // [2] TRS2 - 5'-acgaac-3'

Table 7 - MERS-CoV open-reading frames (ORF) with unknown TRS predicted by Coronavirus.pro

| mRNA | j (bp) | ORF | proteins (Aa) | | Comments |
|---|---|---|---|---|---|
| a | >25532 | ORF3b | AP3b | 66 | Accessory protein AP3b, IRES translation |
| b | >28570 | ORF8b | AP8b1-b7 | 113/105/90 61/55/53/49/37 | Accessory proteins AP8b1-b7 |
| c | >28990 | ORF14 | AP14a/b | 42/108 | Accessory proteins AP14a/b |

SARS-CoV includes a transcription regulating sequence TRS which is not in MERS virus type, TRS3 (5'-cuaaacgaac-3'), also present in 2019-nCoV. SARS-CoV has several accessory protein (AP3a, AP6, AP7a), which are translated directly from mRNAs. In the case of AP6, although in some studies is named as nsp6, it is not processed by any protease, as other non-structural proteins. Other proteins, as AP11 is only characteristic from SARS-CoV. Furthermore, the shortest mRNA (ORF15), with a length of 263bp, translates no significant proteins and has an unknown functionality. Tables 8 and 9 show open-reading frames (ORF) and proteins (structural, non-structural, accessory and fusion) for SARS identified with Coronavirus.pro.

Table 8 - SARS-CoV open-reading frames (ORF) and proteins identified with Coronavirus.pro

| mRNA-TRS[(a)] | j (bp) | ORF | proteins (Aa) | | Fusion protein (Aa) | | Comments |
|---|---|---|---|---|---|---|---|
| 1-[3] | 63 | ORF1.1 | pp1a | 4383 | - | - | [nsp1-nsp11] |
| | | | pp1ab | 7074 | - | - | [nsp1-nsp10], [nsp12-nsp16] |
| 2-[1] | 3665 | ORF1.2 | AP2 | 50 | - | - | unknown |
| | | | pp2a | 3095 | pp2/nsp3 | 894 | [nsp4-nsp11] |
| | | | pp2ab | 5786 | pp2/nsp3 | 894 | [nsp4-nsp10], [nsp12-nsp16] |
| | | | pp2b | 2628 | pp3/nsp12 | 856 | [nsp13-[nsp16] |
| 3-[1] | 3800 | ORF1.2 | AP2 | 50 | - | - | unknown |
| | | | pp2a | 3095 | pp2/nsp3 | 894 | [nsp4-nsp11] |
| | | | pp2ab | 5786 | pp2/nsp3 | 894 | [nsp4-nsp10], [nsp12-nsp16] |
| | | | pp2b | 2628 | pp3/nsp12 | 856 | [nsp13-nsp16] |
| 4-[3] | 21482 | ORF2b | S1p | 1255 | - | - | Surface glycoprotein Spike precursor |
| | | | S1 | 917 | - | - | Surface glycoprotein Spike |
| | | | S0 | 339 | - | - | S0 protease fragment |
| 5-[1] | 21913 | ORF2a | S2p | 1112 | - | - | Surface glycoprotein Spike precursor |
| | | | S2 | 774 | - | - | Surface glycoprotein Spike |
| | | | S0 | 339 | - | - | S0 protease fragment |
| 6-[2] | 25260 | ORF3a | AP3a | 274 | - | - | Accessory protein AP3a (SARS acsp3) |
| 7-[2] | 26109 | ORF4 | E | 76 | - | - | Envelope protein |
| 8-[3] | 26344 | ORF5 | M | 221 | - | - | Membrane protein |
| 9-[2] | 26913 | ORF6 | AP6 | 63 | - | - | Accessory protein AP6 (SARS nsp6) |
| 10-[2] | 27267 | ORF7a | AP7a | 122 | - | - | Accessory protein AP7a |
| 11-[3] | 27769 | ORF8a | AP8a | 40 | - | - | Accessory protein AP8a |
| 12-[2] | 28106 | ORF9a | N | 422 | - | - | Nucleocapsid phosphoprotein (p9a) |
| 13-[1] | 29489 | ORF15 | - | - | - | - | - |

(a) Transcription regulating sequence: [1] TRS-1 - 5'-cuaaac-3' // [2] TRS2 - 5'-acgaac-3' // [3] TRS3 - 5'-cuaaacgaac-3'

Table 9 - SARS-CoV open-reading frames (ORF) with unknown TRS predicted by Coronavirus.pro

| mRNA | j (bp) | ORF | proteins (Aa) | | Comments |
|---|---|---|---|---|---|
| a | > 25478 | ORF3b | AP3b | 175 | Accessory protein AP3b, IRES translation |
| b | >25640 | ORF3b | AP3b2 | 142 | Accessory protein AP3b2 |
| c | > 27273 | ORF7b | AP7b | 45 | AP7b |
| d | > 27779 | ORF8b | AP8b | 85 | AP8b |
| e | >28120 | ORF9b | AP9b | 98 | Accessory protein AP9b, MA15 ExoN1 |
| f | >28130 | ORF11 | AP11 | 73 | AP11 |
| g | > 28500 | ORF14 | AP14a/a | 71/105 | AP14a/b |

2019-nCoV coronavirus presents relative differences in transcription and subgenomic mRNAs translation with other beta-coronavirus. This virus initiates transcription in 20 TRS sites, transcribing more types of subgenomic mRNAs than SARS/MERS. 2019-nCoV synthesizes around 9 types of fusion proteins, which are remarkably more than SARS/MERS coronavirus, and it is expected than the concentration of non-structural proteins, specially nsp12 (RdRp), nsp3 (PL2pro) and nsp5(3CLpro) is also higher. This fact could be related with the most severe health effects (toxicity) and highest infectivity on host patients than other betacoronavirus.

Furthermore, 2019-nCoV transcribes a mRNA (ORF15), with the shortest length (374bp), with no significant proteins encoded, as also present in SARS-CoV. Finally, 2019-nCoV virus translates another accessory protein, AP12, specific of this virus type.

Table 10 and 11 shows open-reading frames (ORF) and proteins (structural, non-structural, accessory and fusion) identified with Coronavirus.pro software in 2019-nCoV virus.

Table 10 - 2019-nCoV open-reading frames (ORF) and proteins identified with Coronavirus.pro

| mRNA-TRS[(a)] | j (bp) | ORF | proteins (Aa) | | Fusion proteins (Aa) | | Comments |
|---|---|---|---|---|---|---|---|
| 1-[3] | 66 | ORF1.1 | pp1a | 4406 | - | - | [nsp1-nsp11] |
| | | | pp1ab | 7097 | - | - | [nsp1-nsp10], [nsp12-nsp16] |
| 2-[1] | 753 | ORF1.2 | pp2a | 4233 | pp2/nsp1 | 8 | [nsp2-nsp11] |
| | | | pp2ab | 6923 | pp2/nsp1 | 8 | [nsp2-nsp10], [nsp12-nsp16] |
| 3-[1] | 2358 | ORF1.3 | pp3a | 3676 | pp3/nsp2 | 89 | [nsp3-nsp11] |
| | | | pp3ab | 6366 | pp3/nsp2 | 89 | [nsp3-nsp10], [nsp12-nsp16] |
| 4-[1] | 3597 | ORF1.4 | AP2a | 47 | - | - | unknown |
| | | | AP2b | 52 | - | - | |
| | | | pp4a | 3095 | pp4/nsp3 | 893 | [nsp3a-nsp11] |
| | | | pp4ab | 5785 | pp4/nsp3 | 893 | [nsp3a-nsp10], [nsp12-nsp16] |
| 5-[1] | 6936 | ORF1.5 | pp5a | 2153 | pp5/nsp3a | 170 | [nsp3b-nsp11] |
| | | | pp5ab | 4843 | pp5/nsp3a | 170 | [nsp3b-nsp10], [nsp12-nsp16] |
| 6-[1] | 8655 | ORF1.6 | pp6a | 1377 | pp6/nsp4 | 234 | [nsp5-nsp11] |
| | | | pp6ab | 4067 | pp6/nsp4 | 234 | [nsp5-nsp10], [nsp12-nsp16] |
| 7-[1] | 13730 | ORF1.7 | pp7 | 2595 | pp7/nsp12 | 824 | [nsp13-nsp16] |
| 8-[1] | 16049 | ORF1.8 | pp8 | 1807 | pp8/nsp12 | 34 | [nsp13-nsp16] |
| 9-[1] | 18452 | ORF1.9 | pp9 | 1019 | pp9/nsp14 | 375 | [nsp15-nsp16] |
| 10-[1] | 20384 | ORF1.10 | pp10 | 374 | pp10/nsp15 | 76 | nsp16 |
| 11-[3] | 21552 | ORF2 | Sp | 1274 | - | - | Surface glycoprotein spike precursor |
| | | | S | 936 | - | - | Surface glycoprotein spike |
| | | | S0 | 338 | - | - | S0 protease fragment |
| 12-[2] | 25385 | ORF3a | AP3a | 276 | - | - | (SARS ORF3/ORF3a/X1/U274) |
| 13-[2] | 26237 | ORF4 | E | 76 | - | - | Envelope protein |
| 14-[3] | 26469 | ORF5 | M | 223 | - | - | Transmembrane protein |
| | | | M1 | 183 | - | - | M1 protease fragment |
| | | | M2 | 40 | - | - | M2 protease fragment |
| 15-[2] | 27041 | ORF6 | AP6 | 62 | - | - | (SARS ORF6/p6) |
| 16-[2] | 27388 | ORF7a | AP7a | 122 | - | - | (SARS ORF8/U122/X4/ORF7a) |
| 17-[1] | 27644 | ORF7b | AP7b | 43 | - | - | (SARS ORF7b) |
| 18-[3] | 27884 | ORF8 | AP8 | 122 | - | - | (SARS ORF8) |
| 19-[3] | 28256 | ORF9 | N | 420 | - | - | Nucleocapsid phosphoprotein (p9a) |
| 20-[1] | 29530 | ORF15 | - | - | - | - | -- |

(a) Transcription regulating sequence: [1] TRS1 - 5’-cuaaac-3’ // [2] TRS2 - 5’-acgaac-3’ // [3] TRS3 - 5’-cuaaacgaac-3’

Table 11 - 2019-nCoV open-reading frames (ORF) with unknown TRS predicted by Coronavirus.pro

| mRNA | j (bp) | ORF | proteins (Aa) | | Comments |
|---|---|---|---|---|---|
| a | >25405 | ORF3b1 | AP3b1 | 42 | AP3b1 |
| | | ORF3b2 | AP3b2 | 34 | AP3b2 |
| b | >25457 | ORF3b3 | AP3b3 | 58 | AP3b3 |
| | | ORF3b4 | AP3b4 | 152 | AP3b4 |
| c | > 28274 | ORF9b | AP9b | 98 | AP9b |
| d | > 28305 | ORF11 | AP11 | 73 | AP11 |
| e | > 28359 | ORF12 | AP12 | 43 | AP12 |
| f | > 28450 | ORF14 | AP14a/b | 74/187 | AP14a/b |

In general to all of these betacoronaviruses, there are several accessory proteins which expression in vivo and in vitro has not been proved, and therefore its function is still unknown. It is the case of accessory protein AP2 in SARS/MERS and AP2a/b in 2019-nCoV.

4.5 Coronavirus proteins

There are considerable differences between spike glycoproteins. For example, the number of spike glycoproteins is variable with MERS and also between SARS virus types. KY417149 (SARS) virus sequence encodes three spike glycoprotein precursors of different amino acid lengths (S1p, S2p and S3p), which later are processed by virus protease into S1, S2 and S3 spikes, with a common fragment S0. In the case of, NC004718 and AY278488 (SARS), it synthesizes two spike precursors (S1p and S2p), whereas 2019-nCoV and MERS, only one is processed. Spike glycoproteins from the same virus, although having different lengths, are estimated with a 100% identity, as observed from their identity matrices.

In the case of other proteins (N, M and E), it can be observed that this virus is more close related to SARS than to MERS, as also discussed previously. However, it presents also around 10% differences with other SARS, so it could be considered as a different virus type

All these proteins have been aligned with Clustal 1.2 to compare similarities [Higgins 1994, Brown et al 1998] (see Annex A for alignment details).

Table 12 compares structural proteins in these genome sequences of beta-coronaviruses.

Table 12 - Comparison of structural proteins of SARS-CoV, MERS-CoV and 2019-nCoV

```
#
#  Structural proteins length (Aa)
#
                            S0     S1p    S1   S2p    S2   S3p    S3    N   M  E
     1: NC019843 (MERS-CoV)   344    1353  1010     -     -     -     -  414 220 83
     2: MN908947 (2019-nCoV)  338    1274   936     -     -     -     -  420 223 76
     3: NC004718 (SARS-CoV)   338    1256   918  1113  775     -     -  423 222 77
     4: AY278488 (SARS-CoV)   338    1256   918  1113  775     -     -  423 222 77
     5: KY417149 (SARS-CoV)   338    1213   875  1242  904   578   240  423 222 77

#  Spike glycoprotein (S) - Percent Identity  Matrix - created by Clustal2.1

     1: NC019843-S   100.00   27.98   29.60   27.98   29.60   28.27   27.36   26.91   44.77
     2: NC004718-S1   27.98  100.00  100.00   99.67   99.74   70.57   75.98   74.19   92.92
     3: NC004718-S2   29.60  100.00  100.00   99.74   99.74   73.94   78.19   78.19   92.92
     4: AY278488-S1   27.98   99.67   99.74  100.00  100.00   70.79   76.21   74.41   92.92
     5: AY278488-S2   29.60   99.74   99.74  100.00  100.00   74.06   78.32   78.32   92.92
     6: MN908947-S    28.27   70.57   73.94   70.79   74.06  100.00   71.31   70.22   89.17
     7: KY417149-S1   27.36   75.98   78.19   76.21   78.32   71.31  100.00  100.00  100.00
     8: KY417149-S2   26.91   74.19   78.19   74.41   78.32   70.22  100.00  100.00  100.00
     9: KY417149-S3   44.77   92.92   92.92   92.92   92.92   89.17  100.00  100.00  100.00

#  S0 protein - Percent Identity  Matrix - created by Clustal2.1
     1: NC019843 (MERS-CoV)    100.00   41.14   41.74   41.74   42.04
     2: MN908947 (2019-nCoV)    41.14  100.00   94.67   94.67   94.67
     3: NC004718 (SARS-CoV)     41.74   94.67  100.00  100.00   98.22
     4: AY278488 (SARS-CoV)     41.74   94.67  100.00  100.00   98.22
     5: KY417149 (SARS-CoV)     42.04   94.67   98.22   98.22  100.00

#  Nucleocapsid (N) - Percent Identity  Matrix - created by Clustal2.1

     1: NC019843    100.00   48.47   48.09   48.09   48.09
     2: MN908947     48.47  100.00   89.29   89.52   89.52
     3: KY417149     48.09   89.29  100.00   99.76   99.76
     4: NC004718     48.09   89.52   99.76  100.00  100.00
     5: AY278488     48.09   89.52   99.76  100.00  100.00

#  Membrane (M) - Percent Identity  Matrix - created by Clustal2.1

     1: NC019843    100.00   40.00   42.27   42.73   42.73
     2: MN908947     40.00  100.00   88.74   89.64   89.64
     3: KY417149     42.27   88.74  100.00   98.20   98.20
     4: NC004718     42.73   89.64   98.20  100.00  100.00
     5: AY278488     42.73   89.64   98.20  100.00  100.00

#  Envelope (E) - Percent Identity  Matrix - created by Clustal2.1
     1: NC019843    100.00   34.67   34.21   34.21   34.21
     2: MN908947     34.67  100.00   96.05   96.00   96.05
     3: KY417149     34.21   96.05  100.00  100.00  100.00
     4: NC004718     34.21   96.00  100.00  100.00  100.00
     5: AY278488     34.21   96.05  100.00  100.00  100.00

#
#  Non-structural proteins length (Aa)
#
                         nsp1/nsp2/nsp3/nsp3a/nsp3b/nsp4/nsp5/nsp6/nsp7/nsp8a/nsp8b/nsp9/nsp10/nsp11
 1: NC019843 (MERS)        181  672 1361   178   348  508  305  292   84   198     -  110   141    14
 2: MN908947 (2019-nCoV)   181  638 1362   218   340  500  306  290   84    56   141  113   139    14
 3: NC004718 (SARS)        180  639 1385   219   340  500  307  289   84    56   141  113   140    13
 4: AY278488 (SARS)        180  638 1364   218   340  500  306  290   84    56   141  113   139    14
 5: KY417149 (SARS)        180  638 1582     -   340  500  306  290   84    56   141  113   139    14


                         nsp12/nsp13a/nsp13b/nsp14/nsp15/nsp16
 1: NC019843 (MERS)         933    236    362   524   340   306
 2: MN908947 (2019-nCoV)    932    601      -   527   346   299
 3: NC004718 (SARS)         931    601      -   527   346   299
 4: AY278488 (SARS)         932    601      -   527   346   299
 5: KY417149 (SARS)         932    601      -   527   346   299
```

2019-nCoV protein sequences have been compared with SARS/MERS, through a distance estimator, calculated as *id(%) = (1-d/L)x100*, where *d* is the Levenshtein distance between both sequences and *L* is the protein length of the SARS/MERS reference protein sequence. Although there are other distance estimators (Needleman-Wunsch, Smith-Waterman, Damerau-Levenshtein), the Levenshtein distance is an accurate estimator for high similar sequences.

In the case of MERS, no identity has been found in any protein (< 50%). Table 13 compares 2019-nCoV proteins with several virus genome sequences of SARS-CoV. As observed, most of non-structural proteins (nsp1, nsp3b and nsp5 to nsp16), accessory proteins AP7a/b and structural proteins M, N and E have the highest percents of similarity (>70%), proving that this virus is more close related with SARS type than MERS. Glycoprotein spike (S), most of accessory proteins (except AP7a/b, AP11 and AP14a) and non-structured proteins nsp2 to nsp3a and nsp4 have low similarity (< 50 %), proving that those proteins are characteristics of this virus type, and potential targets for specific vaccines and antiviral drugs. The fact that non-structural proteins are similar to SARS, indicates that antiviral drugs could be effective also to this virus.

Table 13 - Comparison of structural, non-structural and accessory proteins of 2019-nCoV with SARS-CoV

| | Protein | NC004718 | KY417149 | AY278488 |
|---|---|---|---|---|
| | | 13-AUG-2018 | 18-DEC-2017 | 01-SEP-2009 |
| | | % Id[(1)] | % Id[(3)] | % Id[(2)] |
| Structural | N | 90.48 | 90.24 | 90.48 |
| | M | 90.58 | 89.24 | 90.58 |
| | M1 | 91.26 | 90.16 | 91.80 |
| | M2 | 82.50 | 85.00 | 85.00 |
| | E | 94.74 | 94.74 | 94.74 |
| | Sp | < 50 % | < 50 % | < 50 % |
| | S | | | |
| | S0 | 94.67 | 94.67 | 94.67 |
| Non-structural | Nsp1 | 83.89 | 85.00 | 84.44 |
| | Nsp2 | < 50 % | < 50 % | < 50 % |
| | Nsp3 | | | |
| | Nsp3a | | | |
| | Nsp3b | 87.65 | 87.94 | 88.24 |
| | Nsp4 | < 50 % | < 50 % | < 50 % |
| | Nsp5 | 95.44 | 95.77 | 95.77 |
| | Nsp6 | 87.54 | 87.20 | 87.20 |
| | Nsp7 | 97.62 | 100 | 98.81 |
| | Nsp8a | 94.64 | 98.21 | 98.21 |
| | Nsp8b | 95.74 | 97.16 | 97.16 |
| | Nsp9 | 95.58 | 97.35 | 97.35 |
| | Nsp10 | 96.43 | 97.14 | 96.43 |
| | Nsp11 | 84.62 | 76.92 | 76.92 |
| | Nsp12 | 96.24 | 96.03 | 96.24 |
| | Nsp13 | 99.50 | 99.50 | 99.67 |
| | Nsp14 | 94.69 | 95.64 | 95.07 |
| | Nsp15 | 88.15 | 88.73 | 88.73 |
| | Nsp16 | 93.31 | 94.31 | 93.31 |
| Accessory | AP2a | < 50 % | < 50% | < 50 % |
| | AP2b | | | |
| | AP3a | | | |
| | AP3b1 | | | |
| | AP3b2 | | | |
| | AP3b3 | | | |
| | AP3b4 | | | |
| | AP6 | | 67.74 | |
| | AP7a | 85.25 | 87.70 | 85.25 |
| | AP7b | 79.55 | 81.82 | 79.55 |
| | AP8 | < 50 % | < 50 % | < 50 % |
| | AP9b | | | |
| | AP11 | 76.71 | 73.97 | 76.71 |
| | AP12 | < 50 % | < 50 % | < 50 % |
| | AP14a | 74.32% | 74.32 | 74.32 |
| | AP14b | < 50 % | < 50 % | < 50 % |

As 3CLpro (nsp5) and RdRp (nsp12) have >90% similarities with SARS, some antiviral drugs, such as protease inhibitors or RNA-dependent RNA polymerase inhibitors could be effective to this virus type. Nevertheless, further comparisons would be required, including other types of estimators.

## 5. Conclusions

Coronavirus.pro software provides an accurate and reliable simulation model of Coronaviruses replication cycles: SARS/MERS/2019-nCoV. The code simulates transcription of subgenomic mRNAs, translation, protease cleavage, protein synthesis and virus assembly, including all fusion proteins.

As a result of the analysis, 2019-nCoV can be identified as a beta-coronavirus type SARS-CoV virus with high confidence, named SARS-CoV2, and it is consistent with other recent research analysis. Similarities have been found in 5’utr and 3’utr regions, protease cleavage sites and amino acid composition of both structural and non-structural proteins [Ceraolo and Giorgi 2020, Gorbalenya et al. 2020, Wu et al 2020] However, there are still differences between both coronavirus (SARS-CoV and 2019-nCoV), as the number of spike precursors and accessory proteins.

Coronavirus.pro is able to identify virus type and family, comparing virus genome and proteins with protein and RNA motifs databases. In this case, 2019-nCoV has been identified as a beta-coronavirus SARS in more than 70% than with MERS. However several differences have been found with SARS/MERS. 2019-nCoV has more transcription regulating sequences (TRS) interspaced in the genome and consequently, is producing more subgenomic mRNAs and more fusion proteins during RdRp transcription, which could explain more severe health effects and infectivity than SARS/MERS.

The software has identified those proteins characteristics of 2019-nCoV: Spike S, AP3a, AP3b, AP8, AP9b, AP12 and AP14b and nsp2/3/3a, with similarity < 50 % with other beta-coronaviruses.

Coronavirus.pro has predicted also some accessory proteins in all beta-coronavirus which have not been previously described, called AP2 in SARS-CoV and MERS-CoV, and AP2a/AP2b in 2019-nCoV, respectively. These proteins are encoded in the same genetic region as PL2pro protease (nsp3) and are translated before ORF1.2 (SARS/MERS) and ORF1.4 (2019-nCoV). If they are expressed in vivo or in vitro is not clearly understood, as they could be part of a leaky scanning/shunting mechanism.

The software has predicted some additional protease cleavage sites, giving place to some hypothetical proteins, as nsp3a↓nsp3b, nsp8a↓nsp8b, nsp13a↓nsp13b, M1↓M2 and N1↓N2↓N3. These cleavage sites must be discussed and supported in detail with other methods.

As a conclusion, Coronavirus.pro (2019-nCoV) is able to identify virus genomes and provides in short times useful results (FASTA files of virus proteins and RNA secondary structures). Future research will be focused in interactions between RNA and protein sequences and intracellular processes, fusion protein synthesis, RNA packaging and virus assembly, as carried out before with HIV virus with Monte Carlo simulations. These results will be applied to develop preventive actions (vaccines), diagnostic methods (real time RT-PCR or ELISA tests), and antiviral drugs (fusion inhibitors, RdRp inhibitors or PL2pro/3CLpro protease inhibitors).

Annex A - Sequence alignment of Spike Glycoprotein

A.1 Spike glycoproteins (S)

```
Reference sequence (1): NC019843-S
Identities normalised by aligned length.
Colored by: identity

                         cov    pid    1 [        .         .         .         .         :         .
1 NC019843-S          100.0% 100.0%     MIHSVFLLMFLLTPTESYVDVGPDSVKSACIEVDIQQTFFDKTWPRPIDVSKADGIIYPQG
2 NC004718-S1          88.1%  24.0%     --MFIFLLFLTL---------TSGSDLDRCTTFDDVQAP----NYTQHTSSMRGVYYPDE
3 NC004718-S2          74.3%  26.5%     ------------------------------------------------------------
4 AY278488-S1          88.1%  24.0%     --MFIFLLFLTL---------TSGSDLDRCTTFDDVQAP----NYTQHTSSMRGVYYPDE
5 AY278488-S2          74.3%  26.5%     ------------------------------------------------------------
6 MN908947-S           89.3%  24.5%     --MFVFLV--LL---------PL--VSSQCVNLTTRTQL----PPAYTNSFTRGVYYPDK
7 KY417149-S1          84.0%  23.4%     --------------------------------------------MDKFSSSRRGVYYNDD
8 KY417149-S2          86.8%  22.8%     --MKVLIVLLCL---------GLVTAQDGCGHISTKPQP----LMDKFSSSRRGVYYNDD
9 KY417149-S3          23.7%  43.5%     ------------------------------------------------------------
  consensus/100%                        ............................................................
  consensus/90%                         ............................................................
  consensus/80%                         ............................................................
  consensus/70%                         ............................................................

                         cov    pid   81          .         1         .         .         .         .
1 NC019843-S          100.0% 100.0%     DHGDMYVYSAGHATGTTPQKLFVANYSQDVKQFANGFVVRIGAAANSTGTVIISPSTSATI
2 NC004718-S1          88.1%  24.0%     NVTG---FHTIN-------HT----FGNPVIPFKDGIYFAATE----------------K
3 NC004718-S2          74.3%  26.5%     ------------------------------------------------------------
4 AY278488-S1          88.1%  24.0%     NVTG---FHTIN-------HT----FDNPVIPFKDGIYFAATE----------------K
5 AY278488-S2          74.3%  26.5%     ------------------------------------------------------------
6 MN908947-S           89.3%  24.5%     NVTW---FHAIHVSGTNGTKR----FDNPVLPFNDGVYFASTE----------------K
7 KY417149-S1          84.0%  23.4%     NLTR---YLSFNMDS-ATKVY----FDNPTLPFGDGIYFAATE----------------K
8 KY417149-S2          86.8%  22.8%     NLTR---YLSFNMDS-ATKVY----FDNPTLPFGDGIYFAATE----------------K
9 KY417149-S3          23.7%  43.5%     ------------------------------------------------------------
  consensus/100%                        ............................................................
  consensus/90%                         ............................................................
  consensus/80%                         ............................................................
  consensus/70%                         ............................................................

                         cov    pid  161          .         .         .         2         .         .
1 NC019843-S          100.0% 100.0%     KMGRFFNHTLVLLPDGCGTLLRAFYCI--LEPRSGNHCPAGNSYTSFATYHTPATDCSDGN
2 NC004718-S1          88.1%  24.0%     ---------VIIINNSTNVVIRACNFELCDNPFFAVSKPMG-------------------
3 NC004718-S2          74.3%  26.5%     ---------------------------------------MG-------------------
4 AY278488-S1          88.1%  24.0%     ---------VIIINNSTNVVIRACNFELCDNPFFAVSKPMG-------------------
5 AY278488-S2          74.3%  26.5%     ---------------------------------------MG-------------------
6 MN908947-S           89.3%  24.5%     ---------LLIVNNATNVVIKVCEFQFCNDPFLGVYYHKN-------------------
7 KY417149-S1          84.0%  23.4%     ---------AIIVNNSTHIIIRVCYFNLCKEPMYAISNEQH-------------------
8 KY417149-S2          86.8%  22.8%     ---------AIIVNNSTHIIIRVCYFNLCKEPMYAISNEQH-------------------
9 KY417149-S3          23.7%  43.5%     ------------------------------------------------------------
  consensus/100%                        ............................................................
  consensus/90%                         ............................................................
  consensus/80%                         ........................................t...................
  consensus/70%                         ........................................t...................

                         cov    pid  241          :         .         .         .         .         3
1 NC019843-S          100.0% 100.0%     TFMYTYNITEDEILEWFGITQTAQ-GVHLFSSRYVDLYGG---------NM-----FQFA
2 NC004718-S1          88.1%  24.0%     TFEYISDAFSLDVSEKSGNFKHLREFVFKNKDGFLYVYKGYQPIDVVRDLPSGFNTLKPIF
3 NC004718-S2          74.3%  26.5%     TFEYISDAFSLDVSEKSGNFKHLREFVFKNKDGFLYVYKGYQPIDVVRDLPSGFNTLKPIF
4 AY278488-S1          88.1%  24.0%     TFEYISDAFSLDVSEKSGNFKHLREFVFKNKDGFLYVYKGYQPIDVVRDLPSGFNTLKPIF
5 AY278488-S2          74.3%  26.5%     TFEYISDAFSLDVSEKSGNFKHLREFVFKNKDGFLYVYKGYQPIDVVRDLPSGFNTLKPIF
6 MN908947-S           89.3%  24.5%     TFEYVSQPFLMDLEGKQGNFKNLREFVFKNIDGYFKIYSKHTPINLVRDLPQGFSALEPLV
7 KY417149-S1          84.0%  23.4%     TYDRVGQSFQLDTAPQTGNFKDLREYVFKNKDGFLSVYNAYSPIDIPRGLPVGFSVLKPIL
8 KY417149-S2          86.8%  22.8%     TYDRVGQSFQLDTAPQTGNFKDLREYVFKNKDGFLSVYNAYSPIDIPRGLPVGFSVLKPIL
9 KY417149-S3          23.7%  43.5%     ------------------------------------------------------------
  consensus/100%                        ............................................................
  consensus/90%                         ............................................................
  consensus/80%                         Ta.hh.p.h..-h....G.hpphp.hVah..stah.lYtt..................hh
  consensus/70%                         Ta-hlupsEphDht.ppGNFKpLREaVFKNpDGal.lYpuapPIslsRsLP.GFssLcPlh
```

```
                        cov    pid   401            .          .          .          .          :          .
1 NC019843-S          100.0% 100.0%       VEQAE-GVECDFSPLLSG-TPPQVYNFKRLVFTNCNYNLTKLLSLFSVNDFTCSQISPAAI
2 NC004718-S1          88.1%  24.0%       VRFPNITNLCPFGEVFNATKFPSVYAWERKKISNCVADYSVLYNSTFFSTFKCYGVSATKI
3 NC004718-S2          74.3%  26.5%       VRFPNITNLCPFGEVFNATKFPSVYAWERKKISNCVADYSVLYNSTFFSTFKCYGVSATKI
4 AY278488-S1          88.1%  24.0%       VRFPNITNLCPFGEVFNATKFPSVYAWERKKISNCVADYSVLYNSTFFSTFKCYGVSATKI
5 AY278488-S2          74.3%  26.5%       VRFPNITNLCPFGEVFNATKFPSVYAWERKKISNCVADYSVLYNSTFFSTFKCYGVSATKI
6 MN908947-S           89.3%  24.5%       VRFPNITNLCPFGEVFNATRFASVYAWNRKRISNCVADYSVLYNSASFSTFKCYGVSPTKI
7 KY417149-S1          84.0%  23.4%       IRFPNITNRCPFDKVFNASRFPNVYAWERTKISDCVADYTVLYNSTSFSTFKCYGVSPSKI
8 KY417149-S2          86.8%  22.8%       IRFPNITNRCPFDKVFNASRFPNVYAWERTKISDCVADYTVLYNSTSFSTFKCYGVSPSKI
9 KY417149-S3          23.7%  43.5%       -------------------------------------------------------------
  consensus/100%                          .............................................................
  consensus/90%                           .............................................................
  consensus/80%                           lc.sp.ss.CsFs.lhsu.p.spVYsapRhhhosCshshohLhs.h.hssFpC.tlSsstl
  consensus/70%                           lRFPNITNhCPFscVFNAo+FPsVYAWcRp+ISsCVADYoVLYNSs.FSTFKCYGVSsoKI
```

```
                        cov    pid   481            .        5            .          .          .          .
1 NC019843-S          100.0% 100.0%       KSDLSVSSAGPISQFNYKQSFSNPTCLILATVPHNLTTITKPLKYSYINKCSRLLSDDRTE
2 NC004718-S1          88.1%  24.0%       VRQIAPGQTGVIADYNYKLPDDFMGCVLAWNTRNIDATSTGNYNYKYRYLRHGKLRPFERD
3 NC004718-S2          74.3%  26.5%       VRQIAPGQTGVIADYNYKLPDDFMGCVLAWNTRNIDATSTGNYNYKYRYLRHGKLRPFERD
4 AY278488-S1          88.1%  24.0%       VRQIAPGQTGVIADYNYKLPDDFMGCVLAWNTRNIDATSTGNYNYKYRYLRHGKLRPFERD
5 AY278488-S2          74.3%  26.5%       VRQIAPGQTGVIADYNYKLPDDFMGCVLAWNTRNIDATSTGNYNYKYRYLRHGKLRPFERD
6 MN908947-S           89.3%  24.5%       VRQIAPGQTGKIADYNYKLPDDFTGCVIAWNSNNLDSKVGGNYNYLYRLFRKSNLKPFERD
7 KY417149-S1          84.0%  23.4%       VRQVAPGETGVIADYNYKLPDDFTGCVIAWNTAKQDQ-----GQYYYRSSRKTKLKPFERD
8 KY417149-S2          86.8%  22.8%       VRQVAPGETGVIADYNYKLPDDFTGCVIAWNTAKQDQ-----GQYYYRSSRKTKLKPFERD
9 KY417149-S3          23.7%  43.5%       -------------------------------------------------------------
  consensus/100%                          .............................................................
  consensus/90%                           .............................................................
  consensus/80%                           hpplusupsG.IupaNYK.s.s..sCllhhss.p..t.....hpY.Yh..ppt.Lps.cp-
  consensus/70%                           VRQlAPGpTGsIADYNYKLPDDFhGCVlAWNotp.Dt.....hpYhYR..R+spL+PFERD
```

```
                        cov    pid   561            .          .          .        6            .          .
1 NC019843-S          100.0% 100.0%       T-VWEDGDYYRKQLSPLEGGGWLVASGSTVAMTEQLQMGFGITVQYGTDTNSVCPKLEFAN
2 NC004718-S1          88.1%  24.0%       NCYWPLND-----------YGFYTTTGIGYQPYRVVVLSFELL----NAPATVCGP-----
3 NC004718-S2          74.3%  26.5%       NCYWPLND-----------YGFYTTTGIGYQPYRVVVLSFELL----NAPATVCGP-----
4 AY278488-S1          88.1%  24.0%       NCYWPLND-----------YGFYTTTGIGYQPYRVVVLSFELL----NAPATVCGP-----
5 AY278488-S2          74.3%  26.5%       NCYWPLND-----------YGFYTTTGIGYQPYRVVVLSFELL----NAPATVCGP-----
6 MN908947-S           89.3%  24.5%       NCYFPLQS-----------YGFQPTNGVGYQPYRVVVLSFELL----HAPATVCGP-----
7 KY417149-S1          84.0%  23.4%       --VRTLST-----------YDFYPTVPIEYQATRVVVLSFELL----NAPATVCGP-----
8 KY417149-S2          86.8%  22.8%       --VRTLST-----------YDFYPTVPIEYQATRVVVLSFELL----NAPATVCGP-----
9 KY417149-S3          23.7%  43.5%       -------------------------------------------------------------
  consensus/100%                          .............................................................
  consensus/90%                           .............................................................
  consensus/80%                           ..hh..ts...........hsa.ssss.tht.hc.l.huFtlh....psssoVCs......
  consensus/70%                           ..hhsLss...........YsFhsTssltYQshRVVVLSFELL....sAPATVCGP.....
```

```
                        cov    pid   641            :          .          .          .          .        7
1 NC019843-S          100.0% 100.0%       SGRGVFQNCTAVGVRQQRFVYDAYQNLVGYYSD--DGNYYCLRACVSVPVSVIYD--KETK
2 NC004718-S1          88.1%  24.0%       TGTGVLTPSSKRFQPFQQFGRDVSDFT-DSVRDPKTSEILDISPCAFGGVSVITPGTNASS
3 NC004718-S2          74.3%  26.5%       TGTGVLTPSSKRFQPFQQFGRDVSDFT-DSVRDPKTSEILDISPCAFGGVSVITPGTNASS
4 AY278488-S1          88.1%  24.0%       TGTGVLTPSSKRFQPFQQFGRDVSDFT-DSVRDPKTSEILDISPCSFGGVSVITPGTNASS
5 AY278488-S2          74.3%  26.5%       TGTGVLTPSSKRFQPFQQFGRDVSDFT-DSVRDPKTSEILDISPCSFGGVSVITPGTNASS
6 MN908947-S           89.3%  24.5%       TGTGVLTESNKKFLPFQQFGRDIADTT-DAVRDPQTLEILDITPCSFGGVSVITPGTNTSN
7 KY417149-S1          84.0%  23.4%       KGTGVLTDSSKRFQSFQQFGRDTSDFT-DSVRDPQTLQILDITPCSFGGVSVITPGTNASS
8 KY417149-S2          86.8%  22.8%       KGTGVLTDSSKRFQSFQQFGRDTSDFT-DSVRDPQTLQILDITPCSFGGVSVITPGTNASS
9 KY417149-S3          23.7%  43.5%       -------------------------------------------------------------
  consensus/100%                          .............................................................
  consensus/90%                           .............................................................
  consensus/80%                           pGpGVhp.ssthh...QpFshDh.p.h.s.hpD..s.phhslpsCs.ssVSVIhs..ptop
  consensus/70%                           pGTGVLTsSoK+F.sFQQFGRDsuDhT.DuVRDPpT.pILDIoPCuFGGVSVITPGTNsSs
```

```
                      cov     pid  801         .         .         .         .         :         .
1 NC019843-S        100.0% 100.0%      IQV-DQLNSSYFKLSIPTNFSFGVTQEYIQTTIQKVTVDCKQYVCNGFQKCEQLLREYGQF
2 NC004718-S1        88.1%  24.0%      LGADSSIAYSNNTIAIPTNFSISITTEVMPVSMAKTSVDCNMYICGDSTECANLLLQYGSF
3 NC004718-S2        74.3%  26.5%      LGADSSIAYSNNTIAIPTNFSISITTEVMPVSMAKTSVDCNMYICGDSTECANLLLQYGSF
4 AY278488-S1        88.1%  24.0%      LGADSSIAYSNNTIAIPTNFSISITTEVMPVSMAKTSVDCNMYICGDSTECANLLLQYGSF
5 AY278488-S2        74.3%  26.5%      LGADSSIAYSNNTIAIPTNFSISITTEVMPVSMAKTSVDCNMYICGDSTECANLLLQYGSF
6 MN908947-S         89.3%  24.5%      LGAENSVAYSNNSIAIPTNFTISVTTEILPVSMTKTSVDCTMYICGDSTECSNLLLQYGSF
7 KY417149-S1        84.0%  23.4%      LGAENSIAYANNSIAIPTNFSISVTTEVMPVSMSKTSVDCTMYICGDSQECSNLLLQYGSF
8 KY417149-S2        86.8%  22.8%      LGAENSIAYANNSIAIPTNFSISVTTEVMPVSMSKTSVDCTMYICGDSQECSNLLLQYGSF
9 KY417149-S3        23.7%  43.5%      LGAENSIAYANNSIAIPTNFSISVTTEVMPVSMSKTSVDCTMYICGDSQECSNLLLQYGSF
  consensus/100%                       lts.spls.u..pluIPTNFohulTpEhh.sohtKsoVDCp.YlCss.pcCtpLLhpYGpF
  consensus/90%                        lts.spls.u..pluIPTNFohulTpEhh.sohtKsoVDCp.YlCss.pcCtpLLhpYGpF
  consensus/80%                        LGA-sSlAYuNNoIAIPTNFSISlTTElhPVSMsKTSVDCsMYICGDSpECuNLLLQYGSF
  consensus/70%                        LGA-sSIAYuNNoIAIPTNFSISlTTEVMPVSMuKTSVDCsMYICGDSpECuNLLLQYGSF

                      cov     pid  881         .         9         .         .         .         .
1 NC019843-S        100.0% 100.0%      RNLFASVKSSQSSPIIPGFGGDFNLTLLEPVSISTGSRSARSAIEDLLFDKVTIADPGYMQ
2 NC004718-S1        88.1%  24.0%      REVFAQVKQMYKTPTLKYFGGF-NFSQILPD---PLKPTKRSFIEDLLFNKVTLADAGFMK
3 NC004718-S2        74.3%  26.5%      REVFAQVKQMYKTPTLKYFGGF-NFSQILPD---PLKPTKRSFIEDLLFNKVTLADAGFMK
4 AY278488-S1        88.1%  24.0%      REVFAQVKQMYKTPTLKYFGGF-NFSQILPD---PLKPTKRSFIEDLLFNKVTLADAGFMK
5 AY278488-S2        74.3%  26.5%      REVFAQVKQMYKTPTLKYFGGF-NFSQILPD---PLKPTKRSFIEDLLFNKVTLADAGFMK
6 MN908947-S         89.3%  24.5%      QEVFAQVKQIYKTPPIKDFGGF-NFSQILPD---PSKPSKRSFIEDLLFNKVTLADAGFIK
7 KY417149-S1        84.0%  23.4%      QEVFAQVKQMYKTPAIKDFGGF-NFSQILPD---PSKPTKRSFIEDLLFNKVTLADAGFMK
8 KY417149-S2        86.8%  22.8%      QEVFAQVKQMYKTPAIKDFGGF-NFSQILPD---PSKPTKRSFIEDLLFNKVTLADAGFMK
9 KY417149-S3        23.7%  43.5%      QEVFAQVKQMYKTPAIKDFGGF-NFSQILPD---PSKPTKRSFIEDLLFNKVTLADAGFMK
  consensus/100%                       pplFApVKp..poP.l..FGG..Nho.l.Ps...s.p.otRShIEDLLFsKVTlADsGahp
  consensus/90%                        pplFApVKp..poP.l..FGG..Nho.l.Ps...s.p.otRShIEDLLFsKVTlADsGahp
  consensus/80%                        pEVFAQVKQhYKTPslK.FGGF.NFSQILPD...P.KPoKRSFIEDLLFNKVTLADAGFMK
  consensus/70%                        pEVFAQVKQMYKTPslK.FGGF.NFSQILPD...P.KPTKRSFIEDLLFNKVTLADAGFMK

                      cov     pid  961         .         .         .         0         .         .
1 NC019843-S        100.0% 100.0%      QYVAGYKVLPPLMDVNMEAAYTSSLLGSIAGVGWTAGLSSFAAIPFAQSIFYRLNGVGITQ
2 NC004718-S1        88.1%  24.0%      QKFNGLTVLPPLLTDDMIAAYTAALVSGTATAGWTFGAGAALQIPFAMQMAYRFNGIGVTQ
3 NC004718-S2        74.3%  26.5%      QKFNGLTVLPPLLTDDMIAAYTAALVSGTATAGWTFGAGAALQIPFAMQMAYRFNGIGVTQ
4 AY278488-S1        88.1%  24.0%      QKFNGLTVLPPLLTDDMIAAYTAALVSGTATAGWTFGAGAALQIPFAMQMAYRFNGIGVTQ
5 AY278488-S2        74.3%  26.5%      QKFNGLTVLPPLLTDDMIAAYTAALVSGTATAGWTFGAGAALQIPFAMQMAYRFNGIGVTQ
6 MN908947-S         89.3%  24.5%      QKFNGLTVLPPLLTDEMIAQYTSALLAGTITSGWTFGAGAALQIPFAMQMAYRFNGIGVTQ
7 KY417149-S1        84.0%  23.4%      QKFNGLTVLPPLLTDDMIAAYTAALVSGTATAGWTFGAGAALQIPFAMQMAYRFNGIGVTQ
8 KY417149-S2        86.8%  22.8%      QKFNGLTVLPPLLTDDMIAAYTAALVSGTATAGWTFGAGAALQIPFAMQMAYRFNGIGVTQ
9 KY417149-S3        23.7%  43.5%      QKFNGLTVLPPLLTDDMIAAYTAALVSGTATAGWTFGAGAALQIPFAMQMAYRFNGIGVTQ
  consensus/100%                       QhhsGhpVLPPLhsspM.AtYTuuLluuhhssGWThGhuuhhtIPFA.phhYRhNGlGlTQ
  consensus/90%                        QhhsGhpVLPPLhsspM.AtYTuuLluuhhssGWThGhuuhhtIPFA.phhYRhNGlGlTQ
  consensus/80%                        QKFNGLTVLPPLLTD-MIAAYTuALluGTATuGWTFGAGAALQIPFAMQMAYRFNGIGVTQ
  consensus/70%                        QKFNGLTVLPPLLTDDMIAAYTAALVSGTATAGWTFGAGAALQIPFAMQMAYRFNGIGVTQ

                      cov     pid 1041    ] 1044
1 NC019843-S        100.0% 100.0%      AMQT
2 NC004718-S1        88.1%  24.0%      QIQE
```

A.2 S0 protein from Spike glycoprotein precursor (Sp)

```
                       cov    pid  1 [        .         .         .         .         :         .
1 NC019843          100.0% 100.0%    GFTTTNEAFQKVQDAVNNNAQALSKLASELSNTFGAISASIGDIIQRLDVLEQDAQIDRLIN
2 KY417149           96.8%  40.1%    SLTTTSTALGKLQDVVNQNAQALNTLVKQLSSNFGAISSVLNDILSRLDKVEAEVQIDRLIT
3 NC004718           96.8%  39.8%    SLTTTSTALGKLQDVVNQNAQALNTLVKQLSSNFGAISSVLNDILSRLDKVEAEVQIDRLIT
4 AY278488           96.8%  39.8%    SLTTTSTALGKLQDVVNQNAQALNTLVKQLSSNFGAISSVLNDILSRLDKVEAEVQIDRLIT
5 MN908947           96.8%  39.3%    SLSSTASALGKLQDVVNQNAQALNTLVKQLSSNFGAISSVLNDILSRLDKVEAEVQIDRLIT
  consensus/100%                     uhooTspAhtKlQDsVNpNAQALspLsppLSssFGAISuslsDIlpRLDhlEt-sQIDRLIs
  consensus/90%                      uhooTspAhtKlQDsVNpNAQALspLsppLSssFGAISuslsDIlpRLDhlEt-sQIDRLIs
  consensus/80%                      SLTTTuoALGKLQDVVNQNAQALNTLVKQLSSNFGAISSVLNDILSRLDKVEAEVQIDRLIT
  consensus/70%                      SLTTTuoALGKLQDVVNQNAQALNTLVKQLSSNFGAISSVLNDILSRLDKVEAEVQIDRLIT
```

```
                       cov    pid 81          .         1         .         .         .         .
1 NC019843          100.0% 100.0%    SAALSAQLAKDKVNECVKAQSKRSGFCGQGTHIVSFVVNAPNGLYFMHVGYYPSNHIEVVSA
2 KY417149           96.8%  40.1%    EIRASANLAATKMSECVLGQSKRVDFCGRGYHLMSFPQAAPHGVVFLHVTYVPSQEKNFTTA
3 NC004718           96.8%  39.8%    EIRASANLAATKMSECVLGQSKRVDFCGKGYHLMSFPQAAPHGVVFLHVTYVPSQERNFTTA
4 AY278488           96.8%  39.8%    EIRASANLAATKMSECVLGQSKRVDFCGKGYHLMSFPQAAPHGVVFLHVTYVPSQERNFTTA
5 MN908947           96.8%  39.3%    EIRASANLAATKMSECVLGQSKRVDFCGKGYHLMSFPQSAPHGVVFLHVTYVPAQEKNFTTA
  consensus/100%                     phthSApLAtsKhsECVhuQSKRssFCGpGhHlhSFs.sAPpGlhFhHVsYhPupchphsoA
  consensus/90%                      phthSApLAtsKhsECVhuQSKRssFCGpGhHlhSFs.sAPpGlhFhHVsYhPupchphsoA
  consensus/80%                      EIRASANLAATKMSECVLGQSKRVDFCG+GYHLMSFPQuAPHGVVFLHVTYVPSQE+NFTTA
  consensus/70%                      EIRASANLAATKMSECVLGQSKRVDFCG+GYHLMSFPQuAPHGVVFLHVTYVPSQE+NFTTA
```

```
                       cov    pid 161         .         .         .         2         .         .
1 NC019843          100.0% 100.0%    YFIKTNNTRIVDEWSYTGSSFYAPEPITSLNTKYVAPQVT-YQNISTNLPPPLLGNSTGIDF
2 KY417149           96.8%  40.1%    VFVSN-----GTSWFITQRNFYSPQIITTDNTFVAGNCDVVIGIINNTVYDPLQ--PELDSF
3 NC004718           96.8%  39.8%    VFVFN-----GTSWFITQRNFFSPQIITTDNTFVSGNCDVVIGIINNTVYDPLQ--PELDSF
4 AY278488           96.8%  39.8%    VFVFN-----GTSWFITQRNFFSPQIITTDNTFVSGNCDVVIGIINNTVYDPLQ--PELDSF
5 MN908947           96.8%  39.3%    VFVSN-----GTHWFVTQRNFYEPQIITTDNTFVSGNCDVVIGIVNNTVYDPLQ--PELDSF
  consensus/100%                     hFl.s.....sspW.hTtpsFatPp.ITo.NThhsuspss.ht.lsssl.sPL...sph.sF
  consensus/90%                      hFl.s.....sspW.hTtpsFatPp.ITo.NThhsuspss.ht.lsssl.sPL...sph.sF
  consensus/80%                      VFV.N.....GTpWFlTQRNFauPQIITTDNTFVuGNCDVVIGIINNTVYDPLQ..PELDSF
  consensus/70%                      VFV.N.....GTpWFlTQRNFauPQIITTDNTFVuGNCDVVIGIINNTVYDPLQ..PELDSF
```

```
                       cov    pid 241         :         .         .         .         .         3
1 NC019843          100.0% 100.0%    GSLTQINTTLLDLTYEMLSLQQVVKALNESYIDLKELGNYTYYNKWPWYIWLGFIAGLVALA
2 KY417149           96.8%  40.1%    GDISGINASVVNIQKEIDRLNEVAKNLNESLIDLQELGKYEQYIKWPWYVWLGFIAGLIAIV
3 NC004718           96.8%  39.8%    GDISGINASVVNIQKEIDRLNEVAKNLNESLIDLQELGKYEQYIKWPWYVWLGFIAGLIAIV
4 AY278488           96.8%  39.8%    GDISGINASVVNIQKEIDRLNEVAKNLNESLIDLQELGKYEQYIKWPWYVWLGFIAGLIAIV
5 MN908947           96.8%  39.3%    GDISGINASVVNIQKEIDRLNEVAKNLNESLIDLQELGKYEQYIKWPWYIWLGFIAGLIAIV
  consensus/100%                     GslotINsollslphEh.pLppVsKsLNEShIDLpELGpYp.Y.KWPWYlWLGFIAGLlAls
  consensus/90%                      GslotINsollslphEh.pLppVsKsLNEShIDLpELGpYp.Y.KWPWYlWLGFIAGLlAls
  consensus/80%                      GDISGINASVVNIQKEIDRLNEVAKNLNESLIDLQELGKYEQYIKWPWYlWLGFIAGLIAIV
  consensus/70%                      GDISGINASVVNIQKEIDRLNEVAKNLNESLIDLQELGKYEQYIKWPWYlWLGFIAGLIAIV
```

## A.3 Envelope protein (E)

```
Reference sequence (1): NC019843
Identities normalised by aligned length.
Colored by: identity


                        cov     pid   1 [         .         .         .         .         :         .
1 NC019843           100.0% 100.0%     -LPFVQERIGLFIVNFFIFTVVCAITLLVCMAFLTATRLCVQCMTGFNTLLVQPALYLYN
2 MN908947            91.5%  32.9%     MYSFVSEETGTLIVNSVLLFLAFVVFLLVTLAILTALRLCAYCCNIVNVSLVKPSFYVYS
3 KY417149            92.7%  32.9%     MYSFVSEETGTLIVNSVLLFLAFVVFLLVTLAILTALRLCAYCCNIVNVSLVKPTVYVYS
4 NC004718            92.7%  33.3%     -YSFVSEETGTLIVNSVLLFLAFVVFLLVTLAILTALRLCAYCCNIVNVSLVKPTVYVYS
5 AY278488            92.7%  32.9%     MYSFVSEETGTLIVNSVLLFLAFVVFLLVTLAILTALRLCAYCCNIVNVSLVKPTVYVYS
  consensus/100%                      .hsFVpEchGhhIVN.hlhhlshslhLLVshAhLTAhRLCs.ChshhNs.LVpPshYlYs
  consensus/90%                       .hsFVpEchGhhIVN.hlhhlshslhLLVshAhLTAhRLCs.ChshhNs.LVpPshYlYs
  consensus/80%                       .YSFVSEETGTLIVNSVLLFLAFVVFLLVTLAILTALRLCAYCCNIVNVSLVKPolYVYS
  consensus/70%                       .YSFVSEETGTLIVNSVLLFLAFVVFLLVTLAILTALRLCAYCCNIVNVSLVKPolYVYS

                        cov     pid  81   ] 83
1 NC019843           100.0% 100.0%     WVX
2 MN908947            91.5%  32.9%     ---
3 KY417149            92.7%  32.9%     ---
4 NC004718            92.7%  33.3%     ---
```

MView 1.63, Copyright © 1997-2018 Nigel P. Brown

## A.4 Membrane protein (M)

```
Reference sequence (1): NC019843
Identities normalised by aligned length.
Colored by: identity

                     cov    pid  1 [        .         .         .         .         :         .
1 NC019843        100.0% 100.0%    -MSNMTQLTEAQIIAIIKDWNFAWSLIFLLITIVLQYGYPSRSMTVYVFKMFVLWLLWPSS
2 MN908947        100.0%  39.0%    MADSNGTITVEELKKLLEQWNLVIGFLFLTWICLLQFAYANRNRFLYIIKLIFLWLLWPVT
3 KY417149        100.0%  41.4%    -MAENGTISVEELKRLLEQWNLVIGFLFLAWIMLLQFAYSNRNRFLYIIKLVFLWLLWPVT
4 NC004718        100.0%  41.9%    -MADNGTITVEELKQLLEQWNLVIGFLFLAWIMLLQFAYSNRNRFLYIIKLVFLWLLWPVT
5 AY278488        100.0%  41.9%    -MADNGTITVEELKQLLEQWNLVIGFLFLAWIMLLQFAYSNRNRFLYIIKLVFLWLLWPVT
  consensus/100%                   .hsp.splo.tplhtllcpWNhshuhlFLhhhhlLQauYssRshhlYlhKhhhLWLLWPso
  consensus/90%                    .hsp.splo.tplhtllcpWNhshuhlFLhhhhlLQauYssRshhlYlhKhhhLWLLWPso
  consensus/80%                    .MusNGTITVEELKpLLEQWNLVIGFLFLsWIhLLQFAYuNRNRFLYIIKLlFLWLLWPVT
  consensus/70%                    .MusNGTITVEELKpLLEQWNLVIGFLFLsWIhLLQFAYuNRNRFLYIIKLlFLWLLWPVT

                     cov    pid 81          .         1         .         .         .         .
1 NC019843        100.0% 100.0%    SGIVAAVSAMMWISYFVQSIRLFMRTGSWWSFNPETNCLLNVPFGGTTVVRPLVEDSTSVT
2 MN908947        100.0%  39.0%    AIAMACLVGLMWLSYFIASFRLFARTRSMWSFNPETNILLNVPLHGTILTRPLLESELVIG
3 KY417149        100.0%  41.4%    AIAMACIIGLMWLSYFVASFRLFARTRSMWSFNPETNILLNVPLRGTIVTRPLMESELVIG
4 NC004718        100.0%  41.9%    AIAMACIVGLMWLSYFVASFRLFARTRSMWSFNPETNILLNVPLRGTIVTRPLMESELVIG
5 AY278488        100.0%  41.9%    AIAMACIVGLMWLSYFVASFRLFARTRSMWSFNPETNILLNVPLRGTIVTRPLMESELVIG
  consensus/100%                   uhhhAsl.uhMWlSYFltShRLFhRTtShWSFNPETNhLLNVPhtGThlsRPLhEsphsls
  consensus/90%                    uhhhAsl.uhMWlSYFltShRLFhRTtShWSFNPETNhLLNVPhtGThlsRPLhEsphsls
  consensus/80%                    AIAMACllGLMWLSYFVASFRLFARTRSMWSFNPETNILLNVPL+GTIVTRPLhESELVIG
  consensus/70%                    AIAMACllGLMWLSYFVASFRLFARTRSMWSFNPETNILLNVPL+GTIVTRPLhESELVIG

                     cov    pid161          .         .         .         2         .         .
1 NC019843        100.0% 100.0%    YDRLPNEVTVAKPNVLIALKMVKRQSYGTNSGVAIYHRYKAGNYRSPPITADIE--LALLR
2 MN908947        100.0%  39.0%    IKDLPKEITVATSRTLSYYKLGASQRVAGDSGFAAYSRYRIGNYKLNTDHSSSSDNIALLV
3 KY417149        100.0%  41.4%    IKDLPKEITVATSRTLSYYKLGASQRVGTDSGFAAYNRYRIGNYKLNTDHAGSNDNIALLV
```

## A.5 Nucleocapsid (N)

```
Reference sequence (1): NC019843
Identities normalised by aligned length.
Colored by: identity

                     cov     pid   1 [         .         .         .         .         :         .
1 NC019843         100.0% 100.0%    ------MASPAAPRAVSFADNNDITNTNL----SRGRGRNPKPRAAPNNTVSWYTGLTQHG
2 MN908947          94.7%  43.0%    MSDNGPQ-NQRNAPRITFGGPSDSTGSNQNGERSGARSKQRRPQGLPNNTASWFTALTQHG
3 KY417149          94.9%  42.6%    MSDNGPQSNQRSAPRITFGGPTDSTDNNQNGGRNGTRPKQRRPQGLPNNTASWFTALTQHG
4 NC004718          94.9%  42.6%    MSDNGPQSNQRSAPRITFGGPTDSTDNNQNGGRNGARPKQRRPQGLPNNTASWFTALTQHG
5 AY278488          94.9%  42.6%    MSDNGPQSNQRSAPRITFGGPTDSTDNNQNGGRNGARPKQRRPQGLPNNTASWFTALTQHG
  consensus/100%                    ........s.tss.tloFusssD.TssN.....stsRs+p.+PpuhPNNTsSWaTuLTQHG
  consensus/90%                     ........s.tss.tloFusssD.TssN.....stsRs+p.+PpuhPNNTsSWaTuLTQHG
  consensus/80%                     MSDNGPQuNQRuAPRITFGGPoDSTssNQNGtRsGuRsKQRRPQGLPNNTASWFTALTQHG
  consensus/70%                     MSDNGPQuNQRuAPRITFGGPoDSTssNQNGtRsGuRsKQRRPQGLPNNTASWFTALTQHG

                     cov     pid  81           .         1         .         .         .
1 NC019843         100.0% 100.0%    PAQNAGYWRRQDRKINTGNG-IKQLAPRWYFYYTGTGPEAALPFRAVKDGIVWVHEDGATDA
2 MN908947          94.7%  43.0%    PDDQIGYYRRATRRIRGGDGKMKDLSPRWYFYYLGTGPEAGLPYGANKDGIIWVATEGALNT
3 KY417149          94.9%  42.6%    PDDQIGYYRRATRRVRGGDGKMKELSPRWYFYYLGTGPEASLPYGANKEGIVWVATEGALNT
4 NC004718          94.9%  42.6%    PDDQIGYYRRATRRVRGGDGKMKELSPRWYFYYLGTGPEASLPYGANKEGIVWVATEGALNT
5 AY278488          94.9%  42.6%    PDDQIGYYRRATRRVRGGDGKMKELSPRWYFYYLGTGPEASLPYGANKEGIVWVATEGALNT
  consensus/100%                    PspphGYaRRtsR+lpsGsG.hKpLuPRWYFYYhGTGPEAuLPatAsK-GIlWVtp-GAhss
  consensus/90%                     PspphGYaRRtsR+lpsGsG.hKpLuPRWYFYYhGTGPEAuLPatAsK-GIlWVtp-GAhss
  consensus/80%                     PDDQIGYYRRATRRlRGGDGKMK-LSPRWYFYYLGTGPEAuLPYGANK-GIVWVATEGALNT
  consensus/70%                     PDDQIGYYRRATRRlRGGDGKMK-LSPRWYFYYLGTGPEAuLPYGANK-GIVWVATEGALNT

                     cov     pid 161           .         .         .         2         .         .
1 NC019843         100.0% 100.0%    QFAPGTKLPKNFHIEGTGGNSQSSSRASSLSRNSSRSSSQGSRSGNSTRGTSPGPSGIGAVG
2 MN908947          94.7%  43.0%    QLPQGTTLPKGFYAEGSRGGSQASSRSSSRSRNSSRNSTPGSSRGTSPARMA---GNGGDAA
3 KY417149          94.9%  42.6%    QLPQGTTLPKGFYAEGSRGGSQASSRSSSRSRGNSRNSTPGSSRGNSPARMA---SGGGETA
4 NC004718          94.9%  42.6%    QLPQGTTLPKGFYAEGSRGGSQASSRSSSRSRGNSRNSTPGSSRGNSPARMA---SGGGETA
5 AY278488          94.9%  42.6%    QLPQGTTLPKGFYAEGSRGGSQASSRSSSRSRGNSRNSTPGSSRGNSPARMA---SGGGETA
  consensus/100%                    Qhs.GTpLPKsFahEGotGsSQuSSRuSShSRssSRsSo.GSppGsSstthu...ushGtsu
  consensus/90%                     Qhs.GTpLPKsFahEGotGsSQuSSRuSShSRssSRsSo.GSppGsSstthu...ushGtsu
  consensus/80%                     QLPQGTTLPKGFYAEGSRGGSQASSRSSSRSRssSRNSTPGSSRGNSPARMA...SGGG-sA
  consensus/70%                     QLPQGTTLPKGFYAEGSRGGSQASSRSSSRSRssSRNSTPGSSRGNSPARMA...SGGG-sA

                     cov     pid 241           :         .         .         .         .         3
1 NC019843         100.0% 100.0%    KVKQSQPKVITKKDAAAAKNKMRHKRTSTKSFNMVQAFGLRGPGDLQGNFGDLQLNKLGTED
2 MN908947          94.7%  43.0%    KGQQQQGQTVTKKSAAEASKKPRQKRTATKAYNVTQAFGRRGPEQTQGNFGDQELIRQGTDY
3 KY417149          94.9%  42.6%    KGQQQQGQTVTKKSAAEASKKPRQKRTATKQYNVTQAFGRRGPEQTQGNFGDQDLIRQGTDY
4 NC004718          94.9%  42.6%    KGQQQQGQTVTKKSAAEASKKPRQKRTATKQYNVTQAFGRRGPEQTQGNFGDQDLIRQGTDY
5 AY278488          94.9%  42.6%    KGQQQQGQTVTKKSAAEASKKPRQKRTATKQYNVTQAFGRRGPEQTQGNFGDQDLIRQGTDY
  consensus/100%                    KspQpQspslTKKsAAtAppK.RpKRTuTKtaNhsQAFGhRGPtphQGNFGD.pL.+.GT-.
  consensus/90%                     KspQpQspslTKKsAAtAppK.RpKRTuTKtaNhsQAFGhRGPtphQGNFGD.pL.+.GT-.
  consensus/80%                     KGQQQQGQTVTKKSAAEASKKPRQKRTATKpYNVTQAFGRRGPEQTQGNFGDQ-LIRQGTDY
  consensus/70%                     KGQQQQGQTVTKKSAAEASKKPRQKRTATKpYNVTQAFGRRGPEQTQGNFGDQ-LIRQGTDY

                     cov     pid 321           .         .         :         .         .         .
1 NC019843         100.0% 100.0%    MSQFKLTHQNNDDHGNPVYFLRYSGAIKLDPKNPNYNKWLELLEQNIDAYKTFPKKEKKQKA
2 MN908947          94.7%  43.0%    MSRIGMEVTP------SGTWLTYTGAIKLDDKDPNFKDQVILLNKHIDAYKTFPPTEPKKDK
3 KY417149          94.9%  42.6%    MSRIGMEVTP------SGTWLTYHGAIKLDDKDPQFKDNVILLNKHIDAYKTFPPTEPKKDK
4 NC004718          94.9%  42.6%    MSRIGMEVTP------SGTWLTYHGAIKLDDKDPQFKDNVILLNKHIDAYKTFPPTEPKKDK
5 AY278488          94.9%  42.6%    MSRIGMEVTP------SGTWLTYHGAIKLDDKDPQFKDNVILLNKHIDAYKTFPPTEPKKDK
  consensus/100%                    MSphthphps......sshaLpYpGAIKLDsKsPpapc.l.LLpppIDAYKTFP.pE.Kpct
```

MView 1.63, Copyright © 1997-2018 Nigel P. Brown